\documentclass[usenatbib]{mn2e}
\usepackage{epsfig}
\usepackage{amsmath}
\usepackage{ulem}
\usepackage{times}

\def\gsim{\mathrel{\raise0.35ex\hbox{$\scriptstyle >$}\kern-0.6em 
\lower0.40ex\hbox{{$\scriptstyle \sim$}}}}
\def\lsim{\mathrel{\raise0.35ex\hbox{$\scriptstyle <$}\kern-0.6em 
\lower0.40ex\hbox{{$\scriptstyle \sim$}}}}
\def\halpha{{\rm H$\alpha$}}
\def\oii{{\rm [O{\sc ii}]}}

\def\msun{{\rm M}$_{\odot}$}

\date{\today}
\title[Dusty star formation around a $z\sim 0.8$ cluster]
{Mapping dusty star formation in and around a cluster at \boldmath{z = 0.81} by
wide-field imaging with AKARI}

\author[Y. Koyama et al.]{
\parbox[t]{\textwidth}{
Yusei Koyama$^{1}$\thanks{E-mail: koyama@astron.s.u-tokyo.ac.jp},
Tadayuki Kodama$^{2}$,
Kazuhiro Shimasaku$^{1,3}$,
Sadanori Okamura$^{1,3}$, 
Masayuki Tanaka$^{4}$, 
Hyung Mok Lee$^{5}$,
Myunshin Im$^{5}$,
Hideo Matsuhara$^{6}$,
Toshinobu Takagi$^{6}$,
Takehiko Wada$^{6}$ and
Shinki Oyabu$^{6}$
}
\vspace*{6pt}\\
$^{1}$Department of Astronomy, School of Science, The University of Tokyo,
Tokyo 113-0033, Japan\\
$^{2}$National Astronomical Observatory of Japan, Mitaka, Tokyo 181-8588, Japan\\
$^3$Research Center for Early Universe, School of Science, The
University of Tokyo, Tokyo 113-0033, Japan \\
$^4$European Southern Observatory, Karl-Schwarzschild-Str. 2, D-85748, Garching
bei M\"{u}nchen, Germany \\
$^5$Astronomy Program, FPRD, Department of Physics and Astronomy, Seoul
National University, Seoul 151-742, South Korea \\
$^6$Institute of Space and Astronautical Science, Japan Aerospace
Exploration Agency, Kanagawa 229-8510, Japan \\
}
\begin{document}

\maketitle

%------------------------------------------------------------------
\begin{abstract}
We present environmental dependence of dusty star forming activity 
in and around the cluster RXJ1716.4+6708 at $z= 0.81$ based on 
wide-field and multi-wavelength observations with the Prime Focus 
Camera on the Subaru Telescope (Suprime-Cam) and the Infrared Camera 
onboard the AKARI satellite (IRC).
Our optical data shows that the optical colour distribution 
of galaxies starts to dramatically change from blue to red 
at the medium-density 
environment such as cluster outskirts, groups and filaments.  
By combining with the AKARI infrared data, we find that 15~$ \mu$m-detected
galaxies tend to have optical colours between the red sequence
and the blue cloud with a tail into the red sequence, consistent 
with being dusty star forming galaxies.

The spatial distribution of the 15~$\mu$m-detected galaxies 
over $\sim$ 200 arcmin$^2$ around the cluster reveals that
few 15~$\mu$m galaxies are detected in the cluster central region.
This is probably due to the low star forming activity in the cluster
core.
However, interestingly, the fraction of 15~$\mu$m-detected galaxies 
in the medium-density environments is as high as in the
low-density field, despite the fact that the optical colours start 
to change in the medium-density environments. 
Furthermore, we find that 15~$\mu$m-detected galaxies which have 
optically red colours (candidates for dusty red galaxies) 
and galaxies with high specific star formation rates are 
also concentrated in the medium-density environment.  
These results imply that the star forming activity 
in galaxies in groups and filaments is enhanced due to some
environmental effects specific to the medium-density environment
(e.g. galaxy-galaxy interaction), and such a phenomenon is probably
directly connected to the truncation of star forming activity 
in galaxies seen as the dramatic change in optical colours in such
environment.

\end{abstract}
%------------------------------------------------------------------
%%%%%%%%%%%%%%%%%%%%%%%%%%%%%%%%%%%%%%%%%%%%%%%%%%%%%%%%%%%%%%%%%%
% INTRODUCTION
%%%%%%%%%%%%%%%%%%%%%%%%%%%%%%%%%%%%%%%%%%%%%%%%%%%%%%%%%%%%%%%%%%%
\section{Introduction}
\label{sec:intro}

\subsection{Galaxy properties as a function of environment}
Galaxies live in various environments.  Recent redshift surveys have shown
filamentary large scale structures in the local Universe.
In the distant Universe, at least up to $z \lsim 1$, similar filamentary nature
of large scale structures is found around clusters through wide 
field observations of distant clusters (e.g.\ \citealt{kod05}).
There are also some hints that the large scale structure is
present at much higher redshifts up to $z\sim 6$ 
(\citealt{shi03}; \citealt{ouc05}).

Environment must have played an important role in the history
of galaxy evolution since galaxy properties are strongly dependent 
on environment. 
This is first noted by \cite{dre80}, who showed that early-type
galaxies dominate in high density regions while late-type galaxies
tend to live in low density regions. This trend is called 
``morphology--density relation''. After the Dressler's work, this 
interesting trend was confirmed and extended by many authors (e.g. 
\citealt{pos84}; \citealt{dre97}; \citealt{got03d}; \citealt{pos05}).  
However, it is still unclear
what is the key physical process to produce the morphology--density
relation or other environmental dependence of galaxy properties. 

The morphology--density relation can be understood, at least partly,
as a result of morphological transformation of a substantial number
of galaxies when they entered high-density environments.
Many mechanisms to suppress the star forming activity and to contribute 
to the 
morphological transformation have been proposed (see the review 
by \citealt{bos06}).  For example, 
ram-pressure stripping due to the interaction with hot plasma gas
filled in the cluster core (e.g. \citealt{gun72}), is expected 
to be effective in rich cluster cores.  High-speed encounters
between galaxies, which are often called ``galaxy harassment'',
should occur in very high-density environments (e.g.  \citealt{moo96}).
In addition, mergers or galaxy-galaxy interactions should also
contribute to the galaxy transformation (e.g. \citealt{too72}).
Interactions with the cluster potential may also cause a tidal force
when galaxies pass the central region of clusters (e.g. \citealt{byr90}).
Another mechanism that can be effective is the so-called ``strangulation''
(e.g. \citealt{lar80}), which leads to a slow decline in star formation rate
after a galaxy falls into a more massive (i.e. group or cluster) halo. 
Identification of the key processes behind the galaxy transformation is
one of the major remaining issues in galaxy evolution.

It is well known that the fraction of blue star 
forming galaxies in clusters 
increases towards higher redshifts (Butcher-Oemler effect;
\citealt{but84}). Therefore, the transition from blue 
active galaxies to red passive galaxies should be more 
commonly seen in distant clusters. We can expect to see 
directly such truncation in action through observations 
of distant clusters (see also \citealt{got03a}).
However, importantly, it is reported that most of the actions 
take place in the outskirts of clusters rather than in the cluster core.
In fact, some recent studies focus on the galaxy properties in the
surrounding regions of clusters and try to identify the environment
where the truncation of star formation occurs in accreted galaxies
(e.g. \citealt{abr96}; \citealt{bal99}; \citealt{pim02}).  
In such environment, it is reported that passive spirals 
(i.e.\ spiral morphology but no star formation) tend to be 
found (e.g.\ \citealt{got03b}).
\cite{kod01} performed a wide-field imaging of the CL0939 cluster
at $z=0.41$ and discovered that the colour distribution
changes dramatically at the intermediate density environment
which corresponds to groups/filaments. 
A very similar result was reported by \cite{tan05} for the
surrounding regions of higher redshift clusters, CL0016 at $z=0.55$
and RXJ0152 at $z=0.83$.
They suggest that the intermediate density environment
such as groups or filaments around clusters are the very
sites where the truncation of galaxies is taking place.
These pioneer works are really telling us the need for wide-field
observation of distant clusters in order to study the physical mechanisms
that are responsible for the truncation of galaxies from active phase
to passive phase during the course of hierarchical assembly of galaxies
to clusters.

\subsection{Dusty star forming galaxies in the local and distant Universe}
It is well known that red galaxies tend to have little
star forming activity (passive galaxies), 
while blue galaxies have on-going star-forming activity
(star-forming galaxies) at a given redshift. 
However, the classification of 
passive or star-forming galaxies based only on their
optical colours is sometimes highly uncertain.
In fact, \cite{hai08} showed that $\sim 30 \%$ of field red sequence
galaxies selected from optical colour--magnitude diagrams have on-going
star formation activity with EW(H$\alpha$) $>$ 2\AA, using their 
local galaxy samples from Sloan Digital Sky Survey
(SDSS; \citealt{yor00}).  
\cite{dav06} also showed in their SDSS and SWIRE survey 
(\citealt{lon03})
that $\sim 18 \%$ of their samples have red optical
colours and infrared excess at the same time, which
include both AGNs and on-going dusty star-forming galaxies. 
Similar results are shown for cluster red sequence.
For example, \cite{wol05} studied the Abell 901/902 cluster at $z=0.17$, 
and showed that $\sim 22 \%$ of red sequence galaxies 
are dusty red galaxies from the SED fitting using the 
medium-band survey (COMBO17; \citealt{wol01}). 

Instead of using optical colours, optical emission lines 
(e.g. \oii ($\lambda = 3727$\AA) and H$\alpha $
($\lambda = 6563$ \AA)) are often and widely used
to measure star formation rates (\citealt{ken98}).
However, these lines, especially \oii { } lines in the rest-frame
ultra-violet, are attenuated by inter-stellar dust in the galaxies.  
Also, \oii { } strength is affected by AGN contribution, if any, 
and dependent on metallicity as well.
Although \halpha\, line is a much better indicator than \oii { }
in this respect (and is in fact one of the best indicators
of star formation rates among emission lines), it is redshifted
to near-infrared (NIR) regime at $z \gsim 0.5$, where large format
wide-field camera or spectrograph has become available only recently.
Given these difficulties in optical and NIR observations,
infrared (IR) luminosity of a galaxy that can be sensitively obtained by
space telescopes, serves as an ideal measure of star formation rate,
and it is also well calibrated with local starburst galaxies 
(\citealt{ken98}).
Since the total IR luminosity of a galaxy can be estimated through single
broad-band imaging at mid-infrared (MIR) (e.g. \citealt{tak05}),
it is critically important to observe clusters in MIR bands
as well as in the optical/NIR in order to reveal the hidden star formation
activity and hence trace its true star formation history.
This is especially true at high redshifts, because the 
luminous infrared galaxies (LIRGs) 
which have 
$10^{11} L_{\odot} \le L_{\textrm{IR}} (8-1000\mu \textrm{m}) 
\le 10^{12} L_{\odot}$ 
and ultraluminous infrared galaxies (ULIRGs) which have 
$L_{\textrm{IR}} (8-1000\mu \textrm{m}) \ge 10^{12} L_{\odot}$
are more commonly seen in the distant Universe than in the local Universe
(e.g.\ \citealt{san96}; \citealt{lef05}).

\subsection{Infrared observation of galaxy clusters}

Taking these situations into account, wide-field MIR study of
galaxy clusters covering entire structure around the cluster 
is required to investigate the ``true'' environmental dependence of 
star formation activities of galaxies.  
However, until recently, infrared studies of galaxy clusters 
have been conducted mainly with the ISO satellite (\citealt{kes96}),
which have been limited to very inner regions of clusters 
(see the review by \citealt{met05}). 
Even though, some of these studies showed the importance of large 
amount of hidden star formation activity in the cluster environment 
(e.g. \citealt{duc02}).  
The recent advent of the Spitzer MIPS 
(\citealt{rie04}) has enabled us to investigate wider fields of
distant clusters up to $z\sim 0.8$. 
\cite{gea06} observed very wide field (25$' \times$ 25$'$) around 
CL0024 ($z=0.39$) and MS0451 ($z=0.55$) clusters by mosaicing 
MIPS images.  \cite{mar07} and \cite{bai07} observed RXJ0152 and 
MS1054 (both at $z=0.83$), respectively, but their field coverage 
is limited only to cluster central regions.  \cite{bai07} actually 
imply that for rich clusters at $z \sim 0.8$ even wider-field 
infrared studies are needed.  

In the local Universe, it is well established that
star formation activity depends strongly on environment
in the sense that it systematically declines towards higher density
regions (e.g.\ \citealt{gom03}).  
Recently, however, surprising results are reported
where such a relationship between star formation activity and
local galaxy density becomes inverted at $z \sim 1$ 
(\citealt{elb07}; \citealt{coo07}), 
i.e., star formation rate {\it increases} towards higher density
environment at $z\sim1$.  
This is naively expected that one approaches to the formation epoch
of cluster galaxies which is probably systematically skewed to
higher redshifts compared to the formation epoch of field
galaxies.
Therefore, looking back the environmental dependence of star formation
activity in galaxies as a function of redshift is a basic but vital
step towards understanding the environmentally dependent galaxy
formation and evolution.

In this paper, 
we conduct for the first time a panoramic MIR study of a $z \sim 0.8$
cluster, over a very large area including surrounding groups, filaments 
and outer fields.
Throughout this paper, we use $\Omega_M =0.3$, $\Omega_{\Lambda} =0.7$, 
and $H_0 =70$ km s$^{-1}$Mpc$^{-1}$. Magnitudes are all given in the 
AB system, unless otherwise stated.

%%%%%%%%%%%%%%%%%%%%%%%%%%%%%%%%%%%%%%%%%%%%%%%%%%%%%%%%%%%%%%%%%%%%%%%%%%
% DATA
%%%%%%%%%%%%%%%%%%%%%%%%%%%%%%%%%%%%%%%%%%%%%%%%%%%%%%%%%%%%%%%%%%%%%%%%%%
\section{Data}
\label{sec:data}

%%%%%%%%%%%%%%%%%%%%%%%%%%%%%%%%%%%%%%%%%%%%%
% RXJ1716 cluster
%%%%%%%%%%%%%%%%%%%%%%%%%%%%%%%%%%%%%%%%%%%%%
\subsection{RXJ1716 cluster}
\label{subsec:rxj1716_cluster}

The RXJ1716.4+6708 cluster at $z=0.81$ that we study in this paper 
was first discovered in the ROSAT North Ecliptic Pole Survey 
(NEP: \citealt{hen97}).  Optical spectroscopy was performed by 
\cite{gio99} and 37 cluster members were identified.
Using these spectroscopic samples, \cite{gio99} determined the 
velocity dispersion of this cluster to be $\sigma = 
1522 ^{+215} _{-150}$~km s$^{-1}$.  This value is relatively 
large for its rest-frame X-ray luminosity of
$L_{bol} = 13.86 \pm 1.04  \times 10^{44} $erg s$^{-1}$ and 
the temperature $kT = 6.8 ^{+1.0}_{-0.6}$~keV which are 
based on Chandra data (\citealt{ett04}, 
see also \citealt{gio99}, \citealt{vik02}, and \citealt{toz03}).  
Therefore, it is suggested that this cluster is not totally 
virialised yet.  In fact, the brightest cluster galaxy (BCG) of 
this cluster is located on the northwestern edge of the structure
(\citealt{clo98}), which may be linked to the fact that 
the structure of this cluster is still being formed.
The weak-lensing mass of this cluster is estimated to be 
$2.6 \pm 0.9 \times 10^{14} h^{-1} M_{\odot}$ (\citealt{clo98}). 
This is consistent with the estimated mass based on the
the X-ray data in \cite{ett04};
$M_{\rm{tot}} = 4.35 \pm 0.83  \times 10 ^{14} M_{\odot}$. 
As the X-ray properties of this cluster suggest,
this cluster is relatively rich for a cluster at this redshift.

It has been known that this cluster has a small 
subcluster or group to the northeast of the main cluster, and
the morphology of the X-ray image of this cluster  
elongates towards the subcluster (e.g. \citealt{jel05}).  
\cite{koy07} performed wide-field optical imaging of this
cluster and discovered prominent large-scale structures 
penetrating the cluster core and the second group of this 
cluster towards the southwest of the cluster core based on the photometric 
redshift technique (see Section~\ref{subsec:optical_data}).

%%%%%%%%%%%%%%%%%%%%%%%%%%%%%%%%%%%%%%%%%%%%%
% Optical data
%%%%%%%%%%%%%%%%%%%%%%%%%%%%%%%%%%%%%%%%%%%%%
\subsection{Optical data}
\label{subsec:optical_data}

We have been conducting the PISCES project (\citealt{kod05}).
Taking advantage of the wide field coverage of the Prime
Focus Camera on the Subaru Telescope (Suprime-Cam; \citealt{miy02}),
we have investigated various environments around the X-ray detected 
distant clusters at $0.4 \lsim z \lsim 1.3$ (e.g. \citealt{kod05}; 
\citealt{tan05}, 2006, 2007a,b; \citealt{nak05}; \citealt{koy07}).  
As a part of this project, \cite{koy07} conducted 
a deep and wide-field optical study of the RXJ1716 cluster 
with the Suprime-Cam.
We observed this cluster in $VRi'z'$-bands which neatly 
bracket the 4000\AA { }break feature of $z\sim 0.8$ galaxies.
Catalogues were created with $z' \le 24.9$ galaxies, 
which corresponds to 5$\sigma$ detection limits of our
$z'$-band image. 
We select the cluster members by photometric redshift 
technique (phot-$z$) using the code of \cite{kod99}, and showed the 
prominent large-scale structures around the cluster 
(see Figs.~3 and 4 of \citealt{koy07}).  Galaxies with
$0.76 \le z_{phot} \le 0.83$ were used to map out the 
large-scale structures and this member selection 
is used in this paper.
See \cite{koy07} for the summary of our optical data,
data reduction, and the combined colour image of this cluster.

%%%%%%%%%%%%%%%%%%%%%%%%%%%%%%%%%%%%%%%%%%%%%
% Infrared data
%%%%%%%%%%%%%%%%%%%%%%%%%%%%%%%%%%%%%%%%%%%%% 
\subsection{Infrared data}
\label{subsec:infrared_data}
\begin{figure}
   \begin{center}
    \leavevmode
    \rotatebox{0}{\includegraphics[width=8.5cm,height=8.5cm]{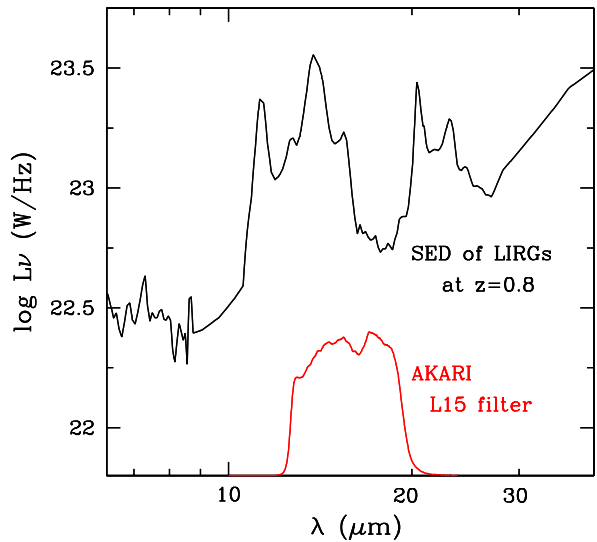}}
   \end{center} 
   \caption{ 
  The response function of the AKARI L15 filter and the template SED
  of starburst galaxies with $10^{11} L_{\odot}$ from Lagache et al. (2004).
  Note that our L15 filter neatly
  captures the PAH broad emissions (7.7~$\mu$m and 8.6~$\mu$m in the 
  rest-frame) from $z\sim 0.8$ galaxies. } 
 \label{fig:L15_filter}
 \end{figure}

We obtained deep and wide-field IR imaging data for 
the RXJ1716 cluster with the Infrared Camera (IRC: \citealt{ona07}) 
onboard the AKARI satellite (\citealt{mur07}). 
Due to the good visibility of NEP directions from AKARI,
we could get very deep data for the RXJ1716 cluster which is 
located near the NEP (see Section \ref{subsec:rxj1716_cluster}). 
The observations were executed from November 26 to December 8 in 2006,
on an open-use programme CLNEP(PI: T. Kodama) and a mission
programme CLEVL(PI: H.M. Lee).
We observed this cluster in the N3(3.3~$\mu$m), 
S7(7.0~$\mu$m) and L15(15.0~$\mu$m) filters, 
using the Astronomical Observation Templates (AOT) 
for deep observation (IRC05) with an AOT filter
combination parameter of ``b'' (see AKARI IRC Data User Manual,
\citealt{lor07}).  We should note that the L15 filter
neatly captures the polycyclic aromatic hydro carbon 
(PAH; \citealt{pug89}) emissions from star-forming
galaxies at $z \sim 0.8$ (Fig.~\ref{fig:L15_filter}).  

%------------------------------------------------------------------
%------------------------------------------------------------------
 \begin{figure*}
    \leavevmode
   \begin{center}
    \leavevmode
    \epsfxsize 0.46\hsize
    \epsfbox{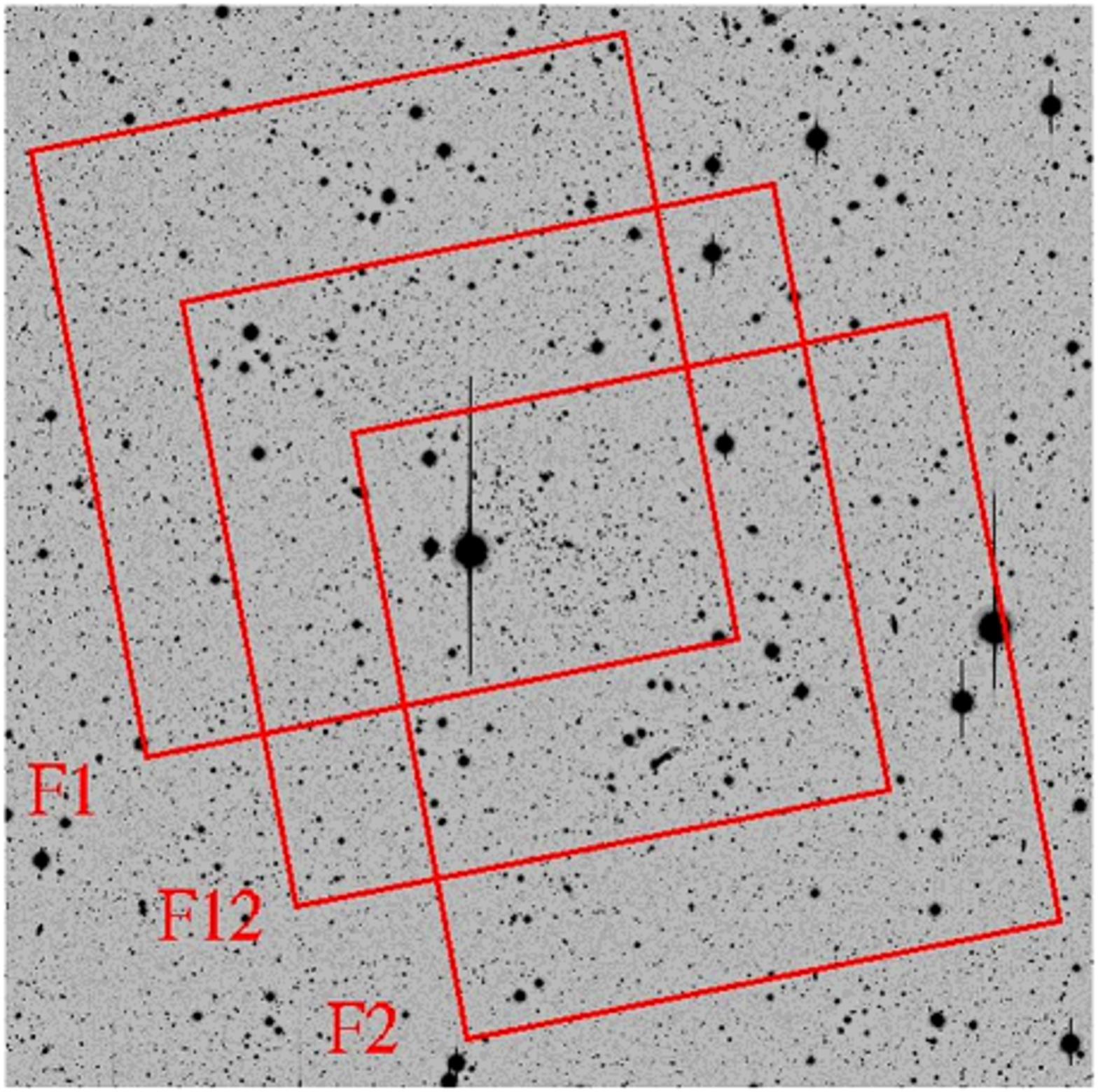}
    \epsfxsize 0.46\hsize
    \epsfbox{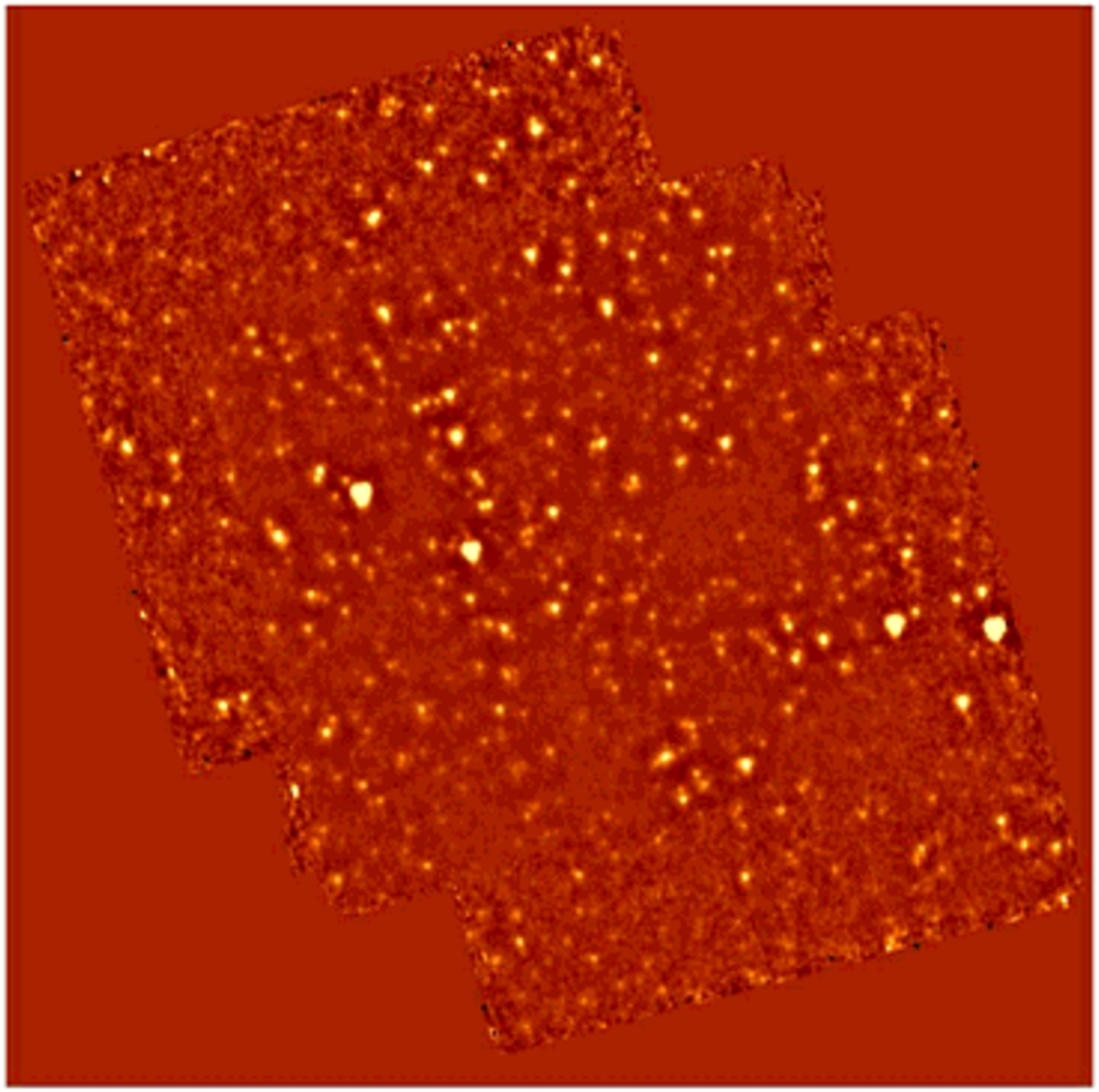}
   \end{center} 
   \caption{ The three AKARI FoVs (F1, F12 and F2) on the $z'$-band
  image ({\it left}) and final co-added L15-band
  image ({\it right}). North is up and east is to
  the left in both of the figures. The size of each 
  panel is $\sim$ 18$'$ $\times$ 18$'$.  }
   \label{fig:L15_image}
 \end{figure*} 
%-----------------------------------------------------------------

We set the three 
field of views (F1, F12 and F2) to cover the large scale structures 
around the RXJ1716 cluster photometrically discovered by \cite{koy07} 
(see Figs.~\ref{fig:L15_image} and \ref{fig:colour_density}b). 
The number of the pointed observations and corresponding
integration time for each field are summarised in 
Table \ref{tab:AKARI_exposure}.
Note that 10 and 30 frames with effective exposure times 44.4 
and 16.4 seconds were included in a single pointed observation
for N3 and S7/L15, respectively.
Therefore, typically, one pointed observation corresponds to
7.4 min and 8.2 min exposure observation for N3 and S7/L15, 
respectively.
IRC has a wide field
of view (about $10' \times 10'$), and our observed fields 
F1, F12 and F2 covered about 200 arcmin$^2$ in total 
(see Figs.~\ref{fig:L15_image} and \ref{fig:colour_density}b). 
We stress here that our study is the first attempt
to cover such a wide field around the $z \gsim 0.8 $ cluster 
in MIR bands. We can study the cluster core to outskirt regions at once 
and this is critically important to discuss the environmental dependence
of the infrared properties of galaxies in and around 
high-$z$ clusters. 

The data were reduced with the IRC imaging pipeline
(version 20071017, \citealt{ona07}) in a standard manner
(see \citealt{lor07}). 
We ran the pipeline for each of the pointed observational data, 
and we made the reduced image for each of the pointed observations 
at first. In this process, we subtracted the median
filtered sky with $20 \times 20$ kernel for 
individual frame before co-adding the frames.
Note that the pipeline often fails to align L15 frames
because of the small number of firmly detected sources. Therefore,
we used ``coaddLusingS'' option in the pipeline, which uses
the alignment of the S7 frames to align the L15 frames
(see also \citealt{lor07} and \citealt{wad07}).   
After that, we matched the position of all the images 
generated from each pointed observational data using the IRAF
tasks '{\it geomap}' and '{\it geotran}', and we co-added 
all the images together. The final 15~$\mu$m image
is shown in the right panel of Fig. \ref{fig:L15_image}. 

Since the exposure times are not uniform over the 
observed field, limiting flux depends on the position.
However, as we observed with 14 pointings for F1 field in N3
and S7 band and with 15 pointings for F12 field in L15 band
(see Table \ref{tab:AKARI_exposure}),
the depths of the data depend mainly on whether a position is
covered by these deep observations or not. 
We thus divide each image into ``deep''  and ``shallow'' 
regions and estimate the sky noise limits of both
regions by a simple aperture photometry of randomly 
distributed apertures in all the images.  Note that
we do not use areas near very bright sources.
The aperture diameters were set to 8$''$, 10$''$ and 11$''$
for N3, S7 and L15, respectively. 
These correspond to twice the size of FWHM of the point spread
function of each image.  We also used these diameters 
for source extraction (see below).  The 5$\sigma$ limiting 
fluxes of deep and shallow regions for each band is summarised in 
Table \ref{tab:AKARI_limit}. Sources were extracted using 
{\sc sextractor} (\citealt{ber96}).  Using the same size of apertures 
(i.e. twice the size of FWHM of each image),  
we measured {\sc flux\_aper} of sources in each image. 
We made the 3~$\mu$m catalogue 
with {\sc  flux\_aper} $\ge 7.0 \mu \rm{Jy}$ in N3 band, 
the 7~$\mu$m catalogue with {\sc flux\_aper} $\ge 21.9 \mu \rm{Jy}$ 
in S7 band and the 15~$\mu$m catalogue with {\sc flux\_aper}
$\ge 66.5 \mu \rm{Jy}$ in L15 band. 
These 5$\sigma$ detection criteria for the deep 
regions approximately correspond to $\sim 3.5\sigma$ detection for the
shallow regions (Table~\ref{tab:AKARI_limit}).   
Also, we use {\sc flux\_auto} and {\sc mag\_auto} 
as the total fluxes and the total magnitudes of the sources 
as shown in \cite{wad07}. 
We note that all the scientific quantities such as total 
IR luminosities and star formation rates are measurements based on 
these total fluxes 
(see also Section~\ref{subsec:derive_sfr}).

%-------------------------------------------------------------------
\begin{table}
\begin{center}
\vspace{1cm}
%\begin{tabular}{p{4zw}|p{7zw}|p{7zw}|p{6zw}}
\begin{tabular}{p{1.1cm}|p{1.8cm}|p{1.8cm}|p{1.3cm}}
\hline
\hline
filter  & F1 & F12 & F2 \\
\hline
N3  &  14  (104 {\it min}) &  2  (15 {\it min})    & 3  (22 {\it min}) \\
S7  &  14  (115 {\it min}) &  3 (25 {\it min})  & 3 (25 {\it min})  \\
L15 &  3 (25 {\it min})  &  15 (123 {\it min})   &  4  (33 {\it min}) \\
\hline 
\end{tabular}
\end{center}
\vspace{0.5cm}
\caption{ Summary of the AKARI(IR) observation. The number of pointings and
corresponding exposure time for each filter and each field is shown. }
\label{tab:AKARI_exposure}
\end{table}
%------------------------------------------------------------------
%----------------------------------------------------------------
\begin{table}
\begin{center}
\vspace{1cm}
%\begin{tabular}[c]{p{5zw}|p{11zw}|p{8zw}}
%\begin{tabular}{c|c|c}
\begin{tabular}[c]{p{2cm}|p{2.5cm}|p{2cm}}
\hline
\hline
filter  &  deep (5$\sigma$)  &  shallow (5$\sigma$)  \\
\hline
N3  &  7.0 $\mu$Jy  & 11.9 $\mu$Jy   \\
S7  &  21.9 $\mu$Jy &  33.0 $\mu$Jy    \\
L15 &  66.5 $\mu$Jy  &  96.5 $\mu$Jy    \\
\hline
\end{tabular}
\end{center}
\vspace{0.5cm}
\caption{ 5$\sigma$ limiting fluxes of N3, S7 and L15 images. 
Deep regions approximately correspond to F1 field of N3/S7 images 
and F12 field of L15 image.
Shallow regions are defined as the rest of the deep regions (see text). }
\label{tab:AKARI_limit}
\end{table}
%-----------------------------------------------------------------

%%%%%%%%%%%%%%%%%%%%%%%%%%%%%%%%%%%%%%%%%%%%%%%%%%%%%%%%%%%%%%%%%%%%%%%%
% ANALYSIS
%%%%%%%%%%%%%%%%%%%%%%%%%%%%%%%%%%%%%%%%%%%%%%%%%%%%%%%%%%%%%%%%%%%%%%%%
\section{Analysis}
\label{sec:analysis}

%%%%%%%%%%%%%%%%%%%%%%%%%%%%%%%%%%%%%%%%%%%%%
% Cross-ID between optical and IR sources
%%%%%%%%%%%%%%%%%%%%%%%%%%%%%%%%%%%%%%%%%%%%%
\subsection{Cross identification between optical and IR sources}
\label{subsec:cross_id}
We need to cross match between the AKARI IR sources and 
the Subaru optical sources.    
We should be very careful when we identify the 
sources in a crowded region like a galaxy cluster.
We have cross-identified the 15~$\mu$m sources in the optical
($z'$-band) image and created ``15~$\mu$m member'' catalogues
in the following way. 

\begin{figure}
   \begin{center}
    \leavevmode
    \rotatebox{0}{\includegraphics[width=8.5cm,height=4.2cm]{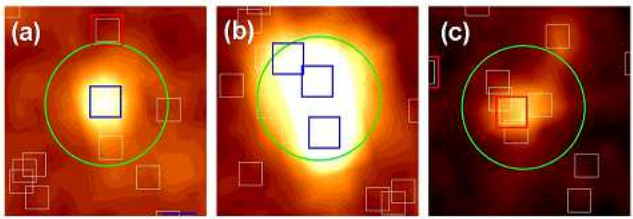}}
   \end{center} 
   \caption{ Examples of the 15~$\mu$m member
  galaxies.  The panel (a) shows an example of ``resolved''
  cluster member while the panels (b) and (c) show
  ``unresolved'' cluster members.  
  In each panel, the large circle indicates the position of the 
  detection of 15~$\mu$m sources, and its radius is 8$''$. Thin-line 
  small squares indicate the positions of the optical sources,
  and the size of them is 3$''$ $\times$ 3$''$.  Thick-line squares
  show the positions of phot-$z$ selected galaxies.
  In the case of the (b) and (c) panels, we select three and 
  one phot-$z$ members as counterparts of the 15~$\mu$m 
  source, respectively.}

\label{fig:ID_example}
 \end{figure}

Firstly, we search for any optical counterpart(s)
using a $8''$ radius from each 15~$\mu$m source.
This radius is sufficiently larger than the FWHM of the PSF in the 
15~$\mu$m image ($\sim$ 5.\hspace{-2pt}$''$5).  The relative
positional accuracy between the $z'$-band image and the L15 image 
is also sufficiently smaller ( $\lsim 1''$) than this search radius. 
We find that 149 of 15~$\mu$m sources have at least one phot-$z$ 
member galaxy (i.e. with $0.76 \le z_{phot} \le 0.83$ as used in 
\cite{koy07} to trace the large scale structure) 
within this radius.  Secondly, we carefully examine
all of them by eye.   
We basically select the nearest object as the optical 
counterpart of each 15~$\mu$m source. 
If the optical counterpart is confirmed to be an 
isolated phot-$z$ member galaxy, we put it in the 
``resolved 15~$\mu$m member catalogue'' (see an example 
in Fig.~\ref{fig:ID_example}a). 
On the other hand, due to the large FWHM of the PSF 
of the 15~$\mu$m image compared to the optical image, 
some 15~$\mu$m sources are not resolved in the source 
extraction process although they show multiple peaks or 
extended shape in the 15~$\mu$m image.  We find that the number 
of such extended sources are not negligible.
Therefore, in such a case, we carefully look at
the shape of the 15~$\mu$m image by eye and judge if the 
15~$\mu$m multi-peak or extended source is associated with 
any phot-$z$ members.  
If we find any phot-$z$ members associated with the
15~$\mu$m source, we put it into the second catalogue, named 
``unresolved 15~$\mu$m member catalogue'' 
(see the examples in Fig.~\ref{fig:ID_example}b,c). 

%------------------------------------------------------------------------
%-------------------------------------------------------------------------

Most of the ``resolved'' 15~$\mu$m members 
have optical counterparts within 2$''$ radius 
(median value is 0.\hspace{-2pt}$''$7), while counterparts of 
``unresolved'' 15~$\mu$m members are found at relatively
larger distance (median value is 2.\hspace{-2pt}$''$5).
We note that in the latter catalogue there are some 
15~$\mu$m sources which have 
more than one optical counterparts (Fig.~\ref{fig:ID_example}b,c). 
For such sources, there is a possibility that the 15~$\mu$m flux of them
is measured as a sum of the fluxes from more than 
one objects.  Therefore, we cannot know its separated individual
15~$\mu$m fluxes correctly and we can only know an upper limit.
The cross identification itself is more difficult for 
unresolved 15~$\mu$m members due to the complexity of 
the shape of the 15~$\mu$m sources.   
Given these problems, we have to treat them carefully 
in the following sections, and in fact we do not include 
these ``unresolved'' 15~$\mu$m members in the
following analyses whenever the 15~$\mu$m fluxes are required.
We count these sources only if detection at 15~$\mu$m is concerned.
However, we have confirmed that the inclusion of such ``unresolved''
sources or not does not affect our results (see also Section~\ 4.4).

For the total of 436 sources detected at 15~$\mu$m in our AKARI field,
we find that 91 of 15~$\mu$m sources are associated with phot-$z$ members. 
In this sample, 54 sources are turned out to be 
resolved 15~$\mu$m members. 
The remaining 37 sources are unresolved cases and we have 65 unresolved
15~$\mu$m members.  This is because about a half of these 37 sources are 
associated with more than one phot-$z$ members. 
Therefore, we have 54 resolved and 65 unresolved 
15~$\mu$m members in total as the numbers of optical counterparts.  
For the resolved 15~$\mu$m members, we search for a counterpart 
in N3 and S7 images in the same way as the cross 
identification between L15 sources and
optical sources described above 
(i.e.\ after searching for N3/S7 counterparts using $8''$ 
radius from each 15~$\mu $m member, checking them by eye ).
In the resolved 15~$\mu$m members, 18 sources are identified
in both N3 and S7 bands, and 24 sources are identified only in
N3 band.  The rest of them are not well distinguished as 
a single object in either N3 or S7 bands.
Our main interest is in the 15~$\mu$m detected cluster members
because these sources are the candidates for dusty star forming 
galaxies.  Therefore, we use 15~$\mu$m detected cluster members
in the following sections regardless of the identification 
in N3 or S7 images. 

Note that photometric redshift is not always the
precise criterion to select the cluster member galaxies.
There remains a possibility that some fore-/background 
contaminations are included in our 15~$\mu $m member catalogues. 
Unfortunately, we cannot estimate the probability 
that a galaxy out of the $0.76\le z_{phot} \le 0.83$ range
falls into this redshift range by our phot-$z$ estimation
because we do not have spectroscopic information
of normal galaxies in our RXJ1716 field.  
However, we do know some spectroscopic redshifts for the X-ray sources 
in our RXJ1716 field from \cite{kim07}.  We find that only 1 out of
27 such sources is miss-identified as our phot-$z$ member.  
Therefore, the fore-/background contamination in our phot-$z$
samples is expected to be very small, although firm conclusion 
waits for a similar check using normal galaxies, which cannot 
be done at the moment because of unavailability of their redshifts.
\cite{mar07}, who studied the RXJ0152 cluster at $z=0.83$, estimated 
that the fraction of 151 spectroscopically confirmed
non-members that fall into the $0.76\le z_{phot} \le 0.88$ range
is only $\sim$ 7\%.  They used the photometric redshifts 
estimated in \cite{tan05} (i.e. phot-$z$ code in 
\citet{kod99}, using the Subaru $VRi'z'$ data), 
which is the same as that used in this study.  
\cite{mar07} also showed that
the photometric redshifts worked well 
for their spectroscopically confirmed MIR member galaxies
($|\Delta z| = |z_{phot} - z_{spec}|\sim  0.01$).    
Therefore, the photometric redshift selection would not do
any critical harm on the statistical properties of our 15~$\mu$m
members, either.
We should, however, confirm the physical association of these
sources to the cluster through spectroscopic follow-up
observations in our future work.

%%%%%%%%%%%%%%%%%%%%%%%%%%%%%%%%%%%%%%%%%%%%%
% Derivation of SFR for 15um members
%%%%%%%%%%%%%%%%%%%%%%%%%%%%%%%%%%%%%%%%%%%%%
\subsection{Derivation of star formation rates for the 15~$\mu$m members}
\label{subsec:derive_sfr}
The AKARI's L15 band corresponds to $\sim 7 - 9$ $\mu$m in the
rest frame of the RXJ1716 cluster at $z=0.81$.  
This wavelength range neatly includes 7.7 and 8.6~$\mu$m 
broad line emission features of PAHs (Fig.~\ref{fig:L15_filter}).
It is well known that MIR broad band luminosity correlates with
total infrared luminosity ($L_{\textrm{IR}}$) and hence 
star formation rate through a good correlation between $L_{\textrm{IR}}$ 
and star formation rate (\citealt{ken98}).
It is particularly true in this case, since the PAH emissions originate
from photo dissociation regions associated to star forming regions and
the intensity of PAH features themselves are in good correlation with
star formation rate (e.g. \citealt{cha01}).  

\begin{figure}
   \begin{center}
    \leavevmode
    \rotatebox{0}{\includegraphics[width=8.5cm,height=8.5cm]{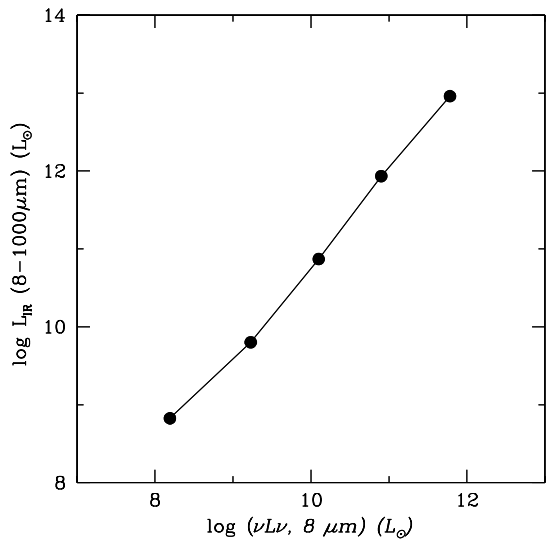}}
   \end{center} 
   \caption{ The relation between $\nu L \nu (8\mu \textrm{m})$ from AKARI 
  data and $L_{\textrm{IR}}(8-1000\mu \textrm{m})$ derived from the template 
  SEDs of Lagache et al. (2004).
  Five points correspond to their five SED templates of starburst galaxies.
  The AKARI 15~$\mu$m flux from a $z=0.81$ galaxy can be converted 
  into a total IR luminosity using this relation (see text).}
\label{fig:nuLnu_vs_Lbol}
 \end{figure}

\begin{figure}
   \begin{center}
    \leavevmode
    \rotatebox{0}{\includegraphics[width=8.5cm,height=8.5cm]{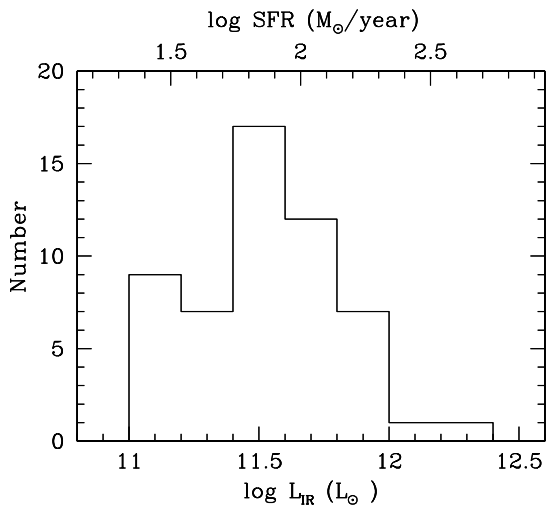}}
   \end{center} 
   \caption{ Distribution of infrared luminosity $L_{\textrm{IR}}$
 (8--1000~$\mu$m) and corresponding SFR for the
  15~$\mu$m cluster members.  Note that $L_{\textrm{IR}}$ and SFRs 
  are calculated only for the resolved 15~$\mu$m members.
\label{fig:histogram_Lbol}}
 \end{figure}

To derive $L_{\textrm{IR}}$ of each galaxy, we use the template 
SEDs of starburst galaxies from \cite{lag04}.
We calculate the conversion factor of our 15~$\mu$m (rest frame
$\sim 8$ $\mu$m) luminosity to the total (8--1000~$\mu$m) 
luminosity using their 5 template SEDs (i.e. templates for
$10^9, 10^{10}, 10^{11}, 10^{12} $ and $10^{13} L_{\odot}$ 
starburst galaxies).  The derived correlation between $\nu L \nu$ (8~$\mu$m)
(AKARI L15 band flux) and $L_{\textrm{IR}}$ (8--1000 $\mu$m) is shown 
in Fig.~\ref{fig:nuLnu_vs_Lbol}.
Using this correlation, we estimate the total IR luminosity
from the measured total 15~$\mu$m flux (i.e. {\sc flux\_auto} value
from {\sc sextractor }) of each 15~$\mu$m member.
Then, using the Kennicutt (1998) calibration:
$$ \textrm{SFR} { } (\textrm{M}_{\odot} \cdot \textrm{yr}^{-1}) =  
4.5 \times 10^{-44} L_{\textrm{IR}} { } (\textrm{erg} \cdot \textrm{s}^{-1}),$$
we derive its star formation rate.
A histogram of $L_{\textrm{IR}}$ and corresponding SFR in our total 
samples are shown in Fig.~\ref{fig:histogram_Lbol}. 
Due to the flux limit of our MIR observation, almost 
all of our MIR samples are turned out to be
LIRGs with $L_{\textrm{IR}} \ge 10^{11} L_{\odot}$. 
Our 15~$\mu$m flux limit of 67~$\mu$Jy corresponds to 
SFR of $\sim$ 25 \msun /yr.  
We calculate $L_{\textrm{IR}}$ and SFR only for the ``resolved'' 
15~$\mu$m members since we cannot separate and measure individual
15~$\mu$m fluxes of the unresolved sources 
(see Section \ref{subsec:cross_id}).
We should note that we do not require accurate absolute
values of $L_{\textrm{IR}}$ and SFR in this paper
and our conclusion will rely rather on their relative
values as a function of environment.

%--------------------------------------------------------------------------
%-------------------------------------------------------------------------
%-------------------------------------------------------------------------
%------------------------------------------------------------------------
 \begin{figure*}
   \begin{center}
    \leavevmode
    \epsfxsize 0.48\hsize
    \epsfbox{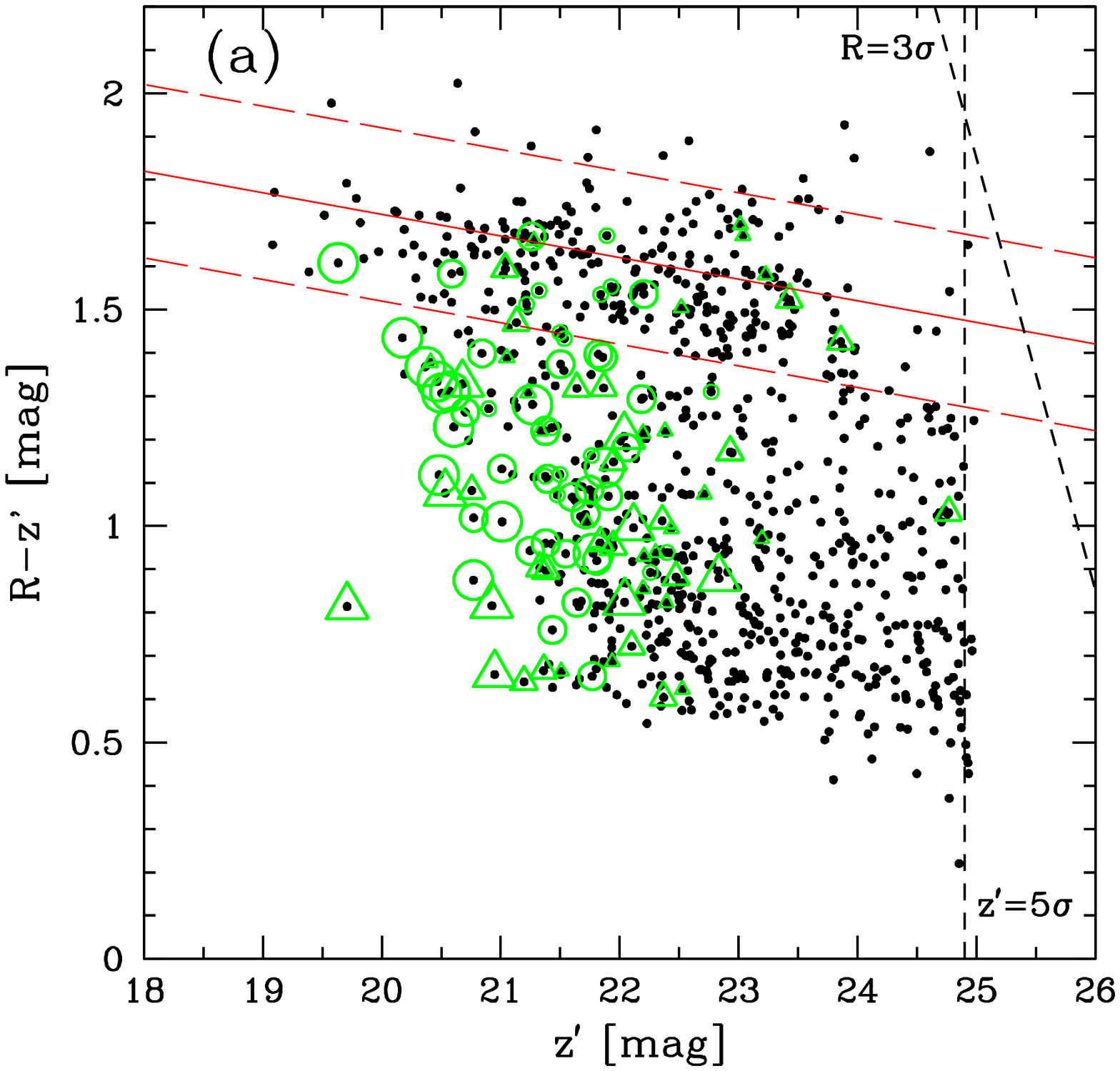}
    \epsfxsize 0.48\hsize
    \epsfbox{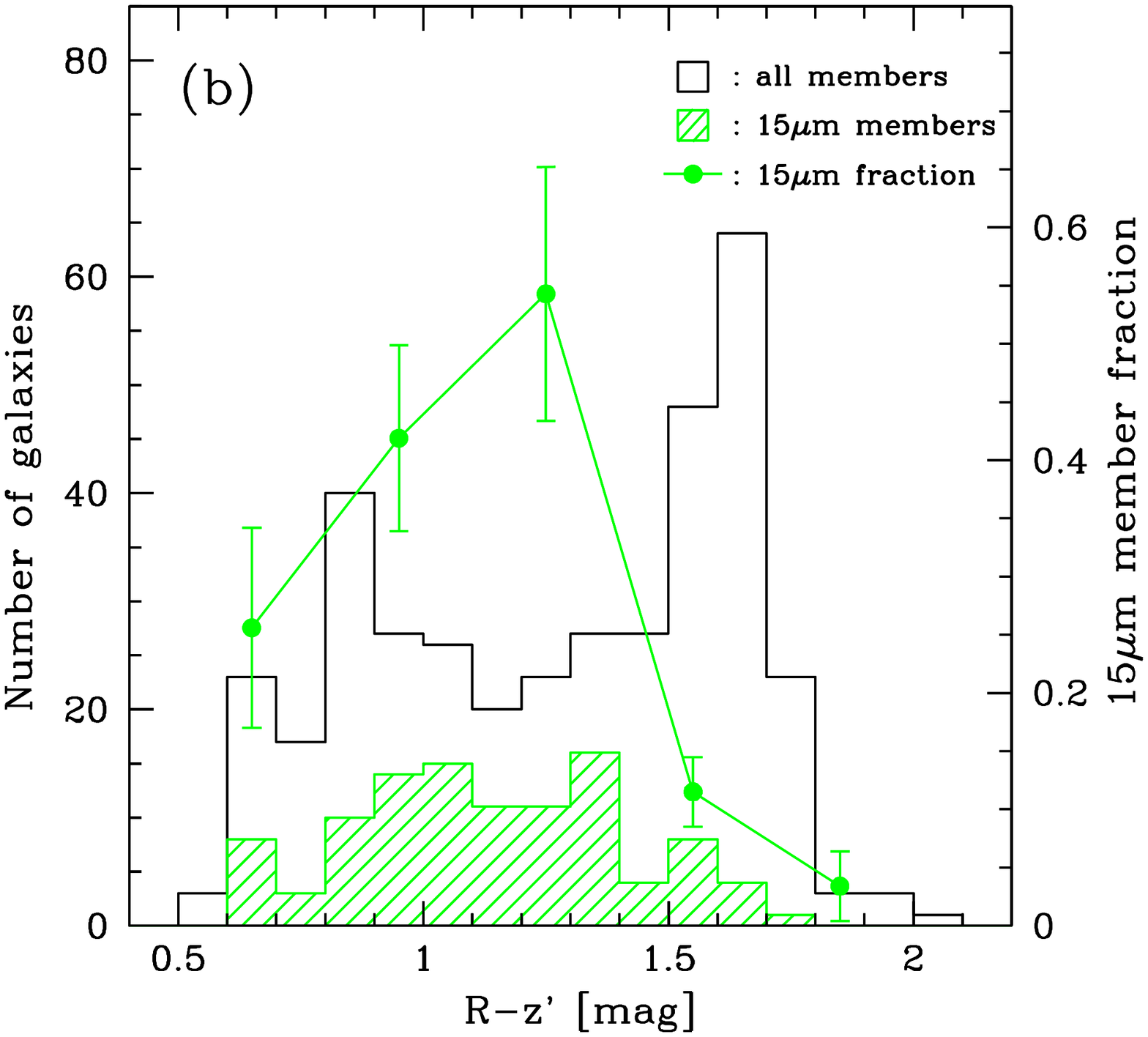}
   \end{center} 
   \vspace{-2.5cm}
   \caption{(a): The colour--magnitude diagram in 
$R-z'$ vs. $z'$ for all the member galaxies in the field 
covered by AKARI FoVs.  Open circles and triangles represent 
the resolved and the unresolved 15~$\mu$m members, respectively.  
The three sizes of the symbols (large, middle and small) indicate
15~$\mu$m flux with ($f(15 \mu m) \ge 160 \mu \textrm{Jy}$, $f(15 \mu 
 m) \ge 100 \mu \textrm{Jy}$ and $f(15 \mu m) \ge 67 \mu \textrm{Jy}$),
respectively. The solid line shows the best-fitted colour--magnitude
relation defined in Koyama et al. (2007), and the two long dashed lines 
indicate $\pm 0.2$mag from the best-fit colour--magnitude relation.
The vertical and slanted short dashed lines show the limiting 
magnitudes in $z'$-band and $R$-band, respectively.
(b): $R-z'$ colour distribution of all the member 
galaxies (open histogram) and the 15~$\mu$m members (hatched histogram).
Histograms are constructed using only the optically bright 
$z' < 22.5$ galaxies in the field covered by AKARI FoVs.  
The solid-line locus indicates the 
fraction of the 15~$\mu$m members in each colour bin
(see the label and the tick marks on the right side).
Error-bars indicate the Poisson errors.  The open histogram shows
two peaks which correspond to red/blue bimodal populations,
while the hatched histogram shows a single broad peak located 
between the two peaks of the open histogram. }
\label{fig:colmag_all}
 \end{figure*}
%-------------------------------------------------------------------------
There is an issue that our IR galaxy samples may contain
contamination from active galactic nuclei (AGNs).
It is reported that the mean AGN fraction in clusters is 
only $\sim 1$\% from optical spectroscopic surveys.
(e.g.\ \citealt{dre99} for their 10 clusters at $ 0.37 < z < 0.56$). 
However, an excess of X-ray detected AGNs is also reported 
in some clusters (e.g. \citealt{mar02}).  
In our 15~$\mu$m member galaxies, we identified three X-ray
point sources using the Chandra X-ray point source catalogue
constructed by \cite{kim07}.  Although we admit that we cannot 
detect all AGNs in X-ray,  these three sources are strong 
candidates for AGNs and we consider that at least a part 
of their 15~$\mu$m fluxes are emitted from AGNs.  
Another possible technique to distinguish AGNs from starburst
galaxies is to use rest-frame NIR colours, based on the
different NIR SED properties of these two populations 
(e.g. \citealt{web06}).  Rest-frame NIR SEDs for 
starburst galaxies are relatively flat, while AGNs
produces power-law SEDs which increases towards 
longer wavelength.  In our data set, the slope of the
NIR SEDs calculated using the fluxes in N3 (rest-frame 1.8~$\mu$m) 
and S7 (rest-frame 3.9~$\mu$m) would be useful.
We therefore investigated the N3$-$S7 colour of
each resolved 15~$\mu$m member galaxy to see if there is
any possible candidate of AGN.
In our 54 resolved 15~$\mu$m members with $f(15\mu m) \ge 67 \mu$Jy, 
18 galaxies are identified in both N3 and S7 bands,
24 are identified only in N3-band, and the remaining 12 are 
identified in neither N3 nor S7 bands 
(see also Section~\ref{subsec:cross_id}).
Out of the 18 galaxies detected in all bands, only two have NIR colour
that is redder than N3$-$S7=0.0, which could be considered as 
AGN candidates.  However, we notice that these two sources 
are not detected in X-ray.

In these ways, we have identified five AGN candidates in total 
from our 15~$\mu$m member galaxies.  
However, we do not exclude these objects from our sample
because we cannot know if all the MIR emissions from these sources
are from AGNs.  It is still possible that starburst is 
also in place in the AGN host galaxies.  
We stress here that the effect of including these AGN candidates 
for our conclusion is negligible (see more detailed discussion in 
Section~\ref{subsec:enhancement_of_dusty_SF}).
We should note, however, that we do not have any
information of NIR SEDs for the objects that are not detected   
in either N3 or S7 bands.
There remains some possibility that some of these objects
could be associated with AGNs, but we cannot exclude them under
present conditions.  
In any case, future spectroscopic follow-up observations of
the 15 $\mu$m members is necessary to discuss the AGN 
contamination further.

%%%%%%%%%%%%%%%%%%%%%%%%%%%%%%%%%%%%%%%%%%%%%%%%%%%%%%%%%%%%%%%%%%%%%%%%
% Results
%%%%%%%%%%%%%%%%%%%%%%%%%%%%%%%%%%%%%%%%%%%%%%%%%%%%%%%%%%%%%%%%%%%%%%%%
\section{Results}
\label{sec:results}

%%%%%%%%%%%%%%%%%%%%%%%%%%%%%%%%%%%%%%%%%%%%%
% 15um members on the CMD
%%%%%%%%%%%%%%%%%%%%%%%%%%%%%%%%%%%%%%%%%%%%%
\subsection{15~$\mu$m members on the colour--magnitude diagram}
\label{subsec:15um_members_on_cmd}

%----------------------------------------------------------------------
 \begin{figure*}
   \begin{center}
    \leavevmode
    \epsfxsize 0.48\hsize
    \epsfbox{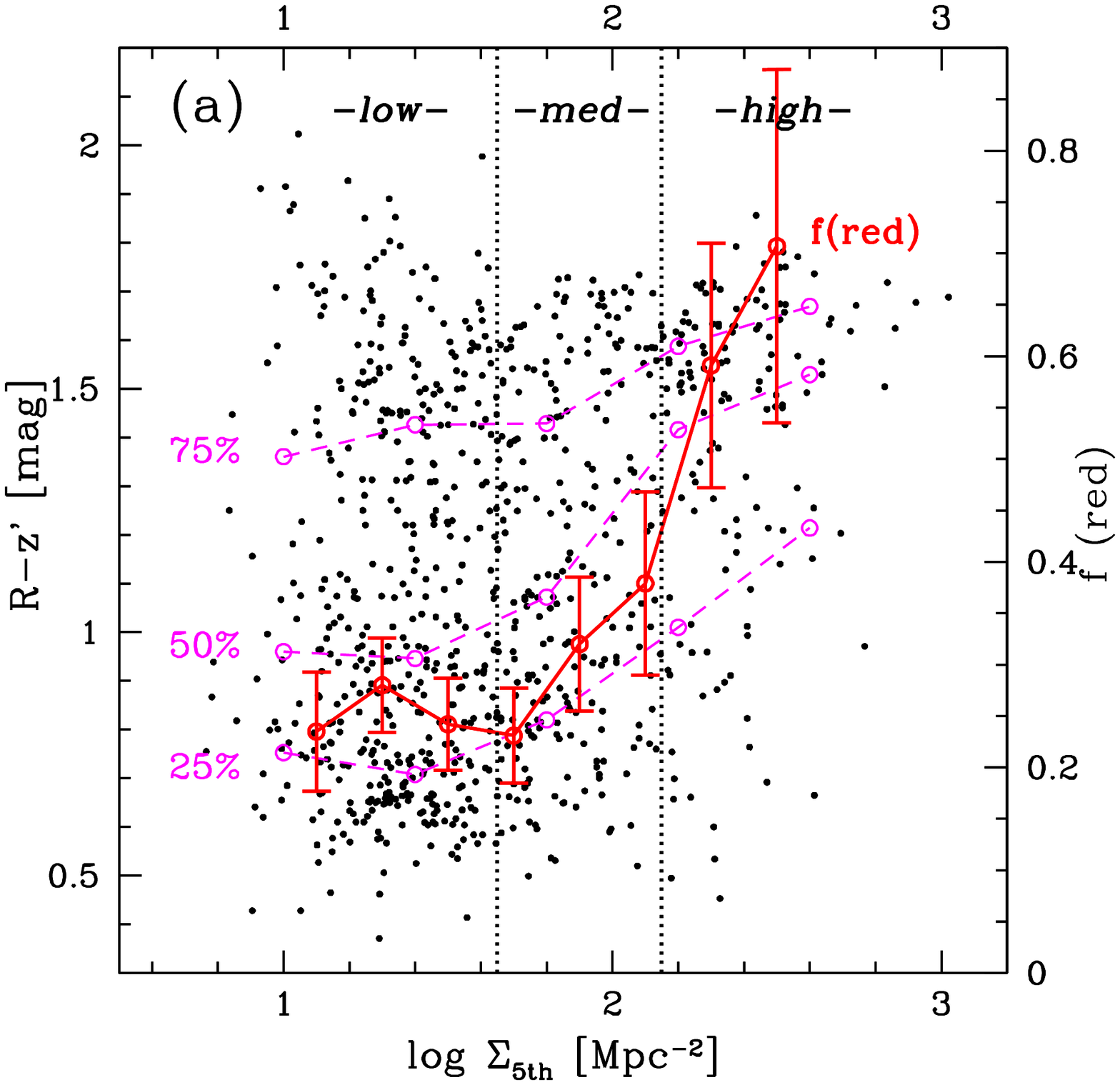}
    \epsfxsize 0.48\hsize
    \epsfbox{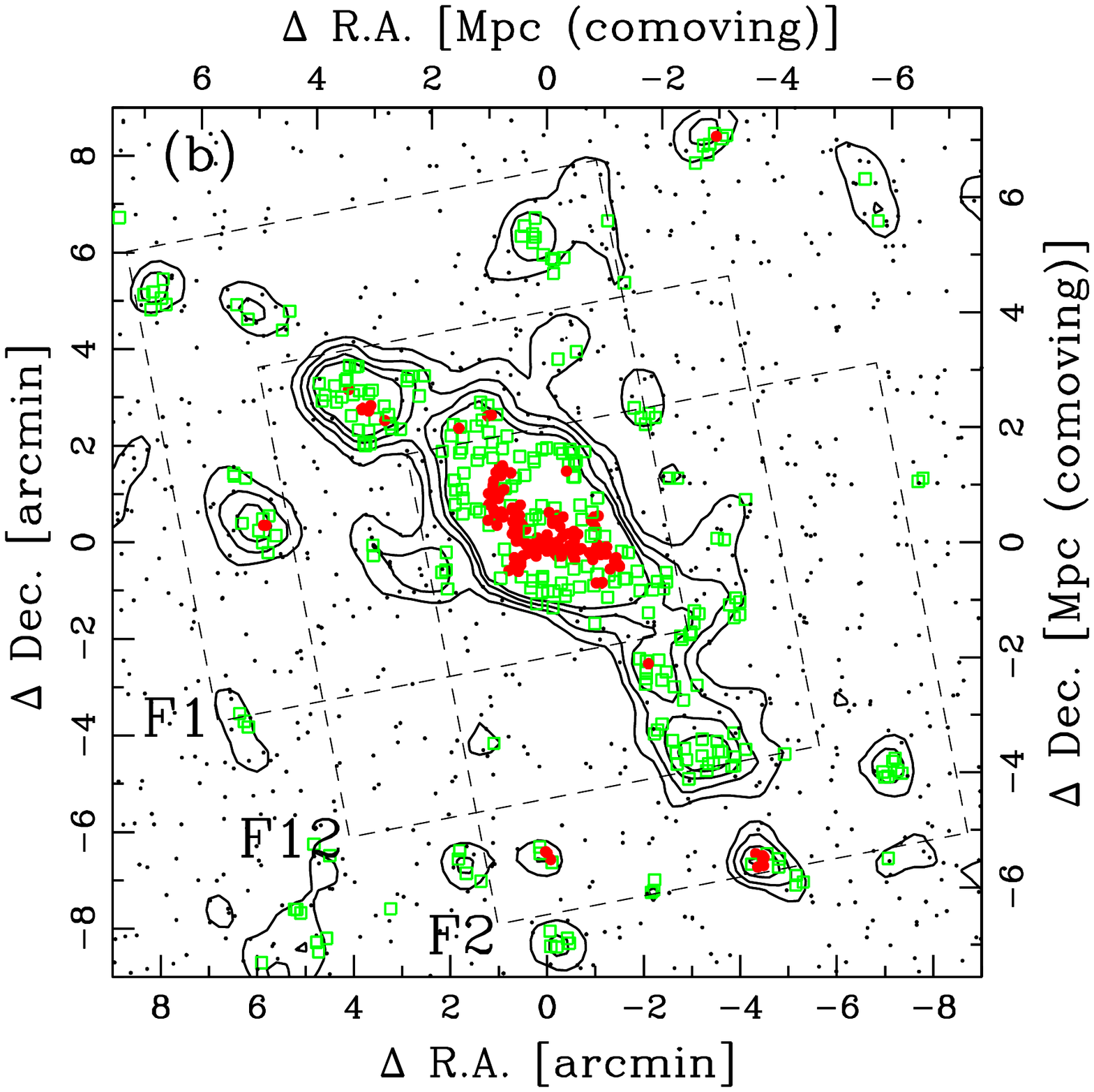}
   \end{center} 
   \vspace{-2.5cm}
\caption{(a): $R-z'$ colours of phot-$z$ member galaxies ($0.76 \le z(phot) 
 \le 0.83$) as a function of local projected number 
 densities of member galaxies within the area covered by our AKARI 
 observations.  
 The vertical dotted lines at $\log \Sigma _{5th}= 1.65$ and
 $\log \Sigma _{5th} = 2.15$ define the ``low'', ``medium'' and ``high''
 density environments as indicated. Three dashed lines represent 
 the loci of the 25th, 50th and 75th percentile colours.  
 The solid line shows the fraction of red galaxies in each 
 environmental bin.  Error bars represent the Poisson errors. 
 (b): Spatial distribution of the phot-$z$ selected 
 galaxies in high/medium/low density environments.  Large filled circles, 
 open squares and small dots represent galaxies in the high, 
 medium and low density environment, respectively.  
 The contours show the local 2D number density of galaxies
 at 2, 3, 4, 5$\sigma$ above the peak of the density distribution.
We apply Gaussian smoothing (sigma= 0.2 Mpc in physical)
on each galaxy and combine the tails of Gaussian wings to 
measure the local density at a given point.  A bin size of
0.1 Mpc (physical) is used to draw iso-density contours.
The three large dashed-line squares show the three 
FoVs of our AKARI pointing observations.
Note that the medium-density environment corresponds to the outskirts
of the cluster core, subclumps and filaments.}
 
\label{fig:colour_density}
\end{figure*}
%-------------------------------------------------------------------------

In this subsection, we present optical properties of
the 15~$\mu$m detected galaxies in the RXJ1716 cluster, 
mainly focusing on optical colours of these galaxies. 
In Fig.~\ref{fig:colmag_all}(a), we show a colour--magnitude diagram
in $R-z'$ vs. $z'$ for the entire field covered by 
the AKARI observations.
All the cluster member galaxies selected on the basis of photometric
redshifts are shown.  The 15~$\mu$m cluster members
are represented by open symbols.  The ``resolved'' and
``unresolved'' 15~$\mu$m members are indicated by open
circles and triangles, respectively.
Most of the 15~$\mu$m members are 
optically bright galaxies with $z' \lsim 22.5$
mainly because of the limited depth of the MIR observations. 

The 15~$\mu$m members are distributed in slightly bluer
side of the red sequence, which is often called a
``green valley'' region.
We show in Fig.~\ref{fig:colmag_all}(b) the histograms of 
$R-z'$ colour distribution for all the phot-$z$ selected 
cluster member galaxies with $z'< 22.5$ (open histogram) 
and for only the 15~$\mu$m member 
galaxies with $z'< 22.5$ (hatched histogram).  
A clear bimodality is seen in the open histogram, while the hatched histogram 
shows a unimodal distribution that fills the gap between the two peaks of
red and blue galaxies.  This trend is quantified by calculating 
the fraction of MIR detected galaxies as a function of $R-z'$ colour 
using the $z'<22.5$ galaxies (a locus in Fig.~\ref{fig:colmag_all}b).  
A similar colour distribution
was reported in \cite{gea06} for two $z\sim 0.5$ clusters 
based on Spitzer MIPS observations.
The medium colours of the MIR detected galaxies bridging the 
``green valley'' would be due to dust reddening of blue 
star forming galaxies.  
We also find that this trend can
be seen even if we limit the sample for relatively bright ($z'<21.5$)
or faint ($21.5 < z' < 22.5$) galaxies.  We should note, however,
that this trend might be produced partly due to the small systematic
difference in luminosity of galaxies between those in the green valley
(brighter) and those in the blue cloud (fainter). 

It is interesting to note that some red-sequence galaxies
are detected at 15~$\mu$m.  It can be interpreted that they are forming
stars but are highly reddened by dust.  Recently, 
it is reported that the fraction of such MIR detected 
star forming galaxies on the red sequence in the total 
star forming galaxies increases towards distant clusters 
(\citealt{sai08}). 
We hereafter call these optically red 15~$\mu$m members
``dusty red galaxies'', and discuss them 
further in Section~\ref{subsec:enhancement_of_dusty_SF}.

%%%%%%%%%%%%%%%%%%%%%%%%%%%%%%%%%%%%%%%%%%%%%
% Optical colour transition in the outskirts
%%%%%%%%%%%%%%%%%%%%%%%%%%%%%%%%%%%%%%%%%%%%%
\subsection{Optical colour transition in the outskirts of the cluster}
\label{subsec:optical_colour_transition}

%---------------------------------------------------------------
 \begin{figure*}
   \begin{center}
    \leavevmode
    \epsfxsize 0.48\hsize
    \epsfbox{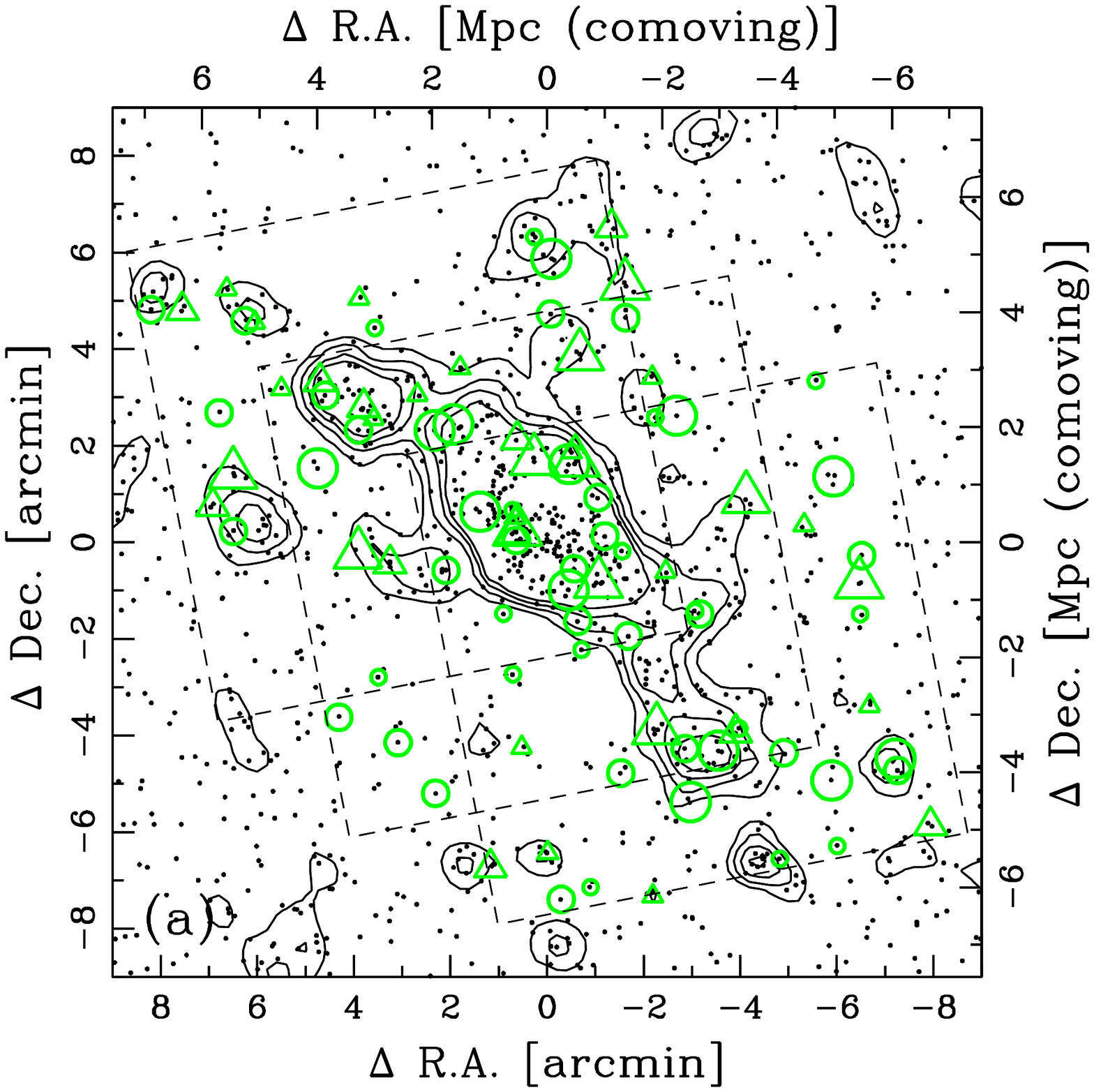}
    \epsfxsize 0.48\hsize
    \epsfbox{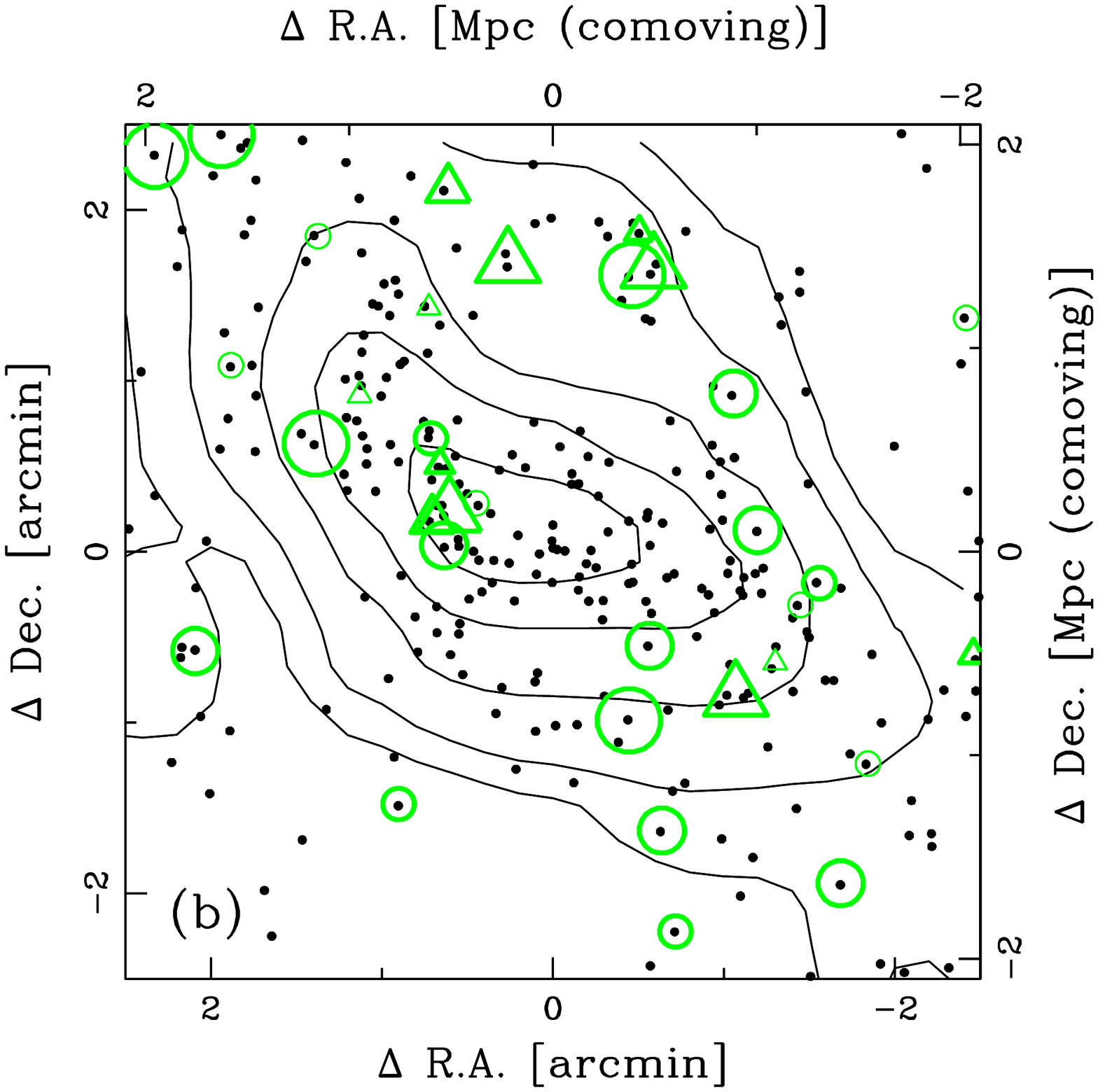}
   \end{center}
   \vspace{-2.5cm}
   \caption{ (a) Spatial distribution of the 15~$\mu$m members around the
cluster for the entire field covered by our AKARI observation. Small 
dots show all the phot-$z$ member galaxies.
Open circles and triangles indicate the resolved and the 
unresolved 15~$\mu$m members, respectively.
The sizes of the open circles and triangles indicate 15~$\mu$m fluxes
as in the previous figure (Fig.~\ref{fig:colmag_all}) . 
Contours are drawn in the same way as in Fig.~\ref{fig:colour_density}b.
The three tilted dashed-line squares show our three AKARI FoVs.
(b): Close-up view of the central $5' \times 5'$ region of the left panel.
The meanings of the symbols are the same as 
in the left panel, but contour levels are changed to
3, 5, 10, 15, 20$\sigma$ above the peak of the density distribution.  
In this plot,
15~$\mu$m members are shown down to 3.5$\sigma$ level as 
thin-line tiny open symbols.   
Note that few 15~$\mu$m members are detected in the very
central region (i.e. $\lsim 1'$ from the cluster centre). }
\label{fig:map_IR}
 \end{figure*}
%---------------------------------------------------------------

In this subsection, we define galaxy environment on the basis of
local projected number density of galaxies.
We use a nearest-neighbour density, which is widely used 
by many authors (e.g. \citealt{tan05}). 
The number density at each galaxy's position is calculated using
the neighbouring galaxies located within the circle of the radius
that is equal to the distance from the galaxy to the 5th-nearest galaxy
(hereafter, $\Sigma _{5th}$).  In this calculation, all the 
optically selected phot-$z$ members are used.

Fig.~\ref{fig:colour_density}(a) shows the $R-z'$ colours 
of individual member galaxies as a function of the above 
defined local density ($\Sigma _{5th}$).
The colour distribution changes dramatically at
$\log \Sigma _{5th} \sim 2.0 $ as the median (50 \%) colour 
locus suggests. 
It can also clearly be seen in
the fraction of red galaxies that starts to
increase sharply at $ \log \Sigma _{5th} \gsim 1.7 $.
We divide the galaxies into three environmental bins,
namely, low-density ($\log \Sigma _{5th} < 1.65$), medium-density
($ 1.65 \le \log \Sigma _{5th} < 2.15 $) and 
high-density ($\log \Sigma _{5th} \ge 2.15 $) regions.  
The medium-density region is defined to a relatively narrow
range of local density where the optical 
colour distribution starts to change.  
These definitions of the environments may seem a little arbitrary, 
but our results do not change if we slightly change
the definitions of the environments. 
In Fig.~\ref{fig:colour_density}(b), we show the spatial 
distribution of the member galaxies in each environment
with different symbols.
The high-density region corresponds approximately to the cluster core
region.  The medium-density environment trace outskirts of the 
cluster core, groups and filaments.
The low-density environment corresponds to the remaining fields 
outside the groups and filaments.   We note that the median value 
of $\log \Sigma_5$ for all the galaxies with $0.76 \le z(phot) \le 0.83$
in our entire Suprime-Cam field (34$' \times$27$'$) is 
$\sim 1.3$.  Excluding the cluster region would only lower 
this value by a negligible amount.  
Therefore, we can reasonably consider that our low-density region
corresponds to general field at the $z\sim 0.8$ Universe at least 
in terms of the local density of galaxies. 

The above finding that the colour distribution dramatically changes
in the ``medium-density'' environment is consistent with 
the previous works.  \cite{kod01} showed similar 
results for the $z\sim 0.4$ cluster, and \cite{tan05}
also found similar trends for $z = 0.55 $ and $z=0.83$ clusters, 
both based on the panoramic imaging with Suprime-Cam on the Subaru
Telescope. 
The current result for another cluster (RXJ1716) lends further support
to the scenario that galaxies start to be truncated in groups or outskirts
of clusters before they entered a very high-density environment such as
cluster cores.
This scenario requires some mechanism that works efficiently in 
relatively low-density environment and truncates star formation activity.

%%%%%%%%%%%%%%%%%%%%%%%%%%%%%%%%%%%%%%%%%%%%%%
% Low star formation rates in the cluster core
%%%%%%%%%%%%%%%%%%%%%%%%%%%%%%%%%%%%%%%%%%%%%%
\subsection{Low star formation rate in the cluster core}
\label{subsec:low_sfr_in_cluster_core}

We show in Fig.~\ref{fig:map_IR}(a) the spatial distribution 
of 15~$\mu$m member galaxies on top of the distribution of 
phot-$z$ selected member galaxies over the $18' \times 18' $ 
area centred on the RXJ1716 cluster.
This area corresponds to $\sim$ 15 Mpc $\times$ 15 Mpc
in comoving scale at $z=0.81$. 
The open circles and triangles indicate the positions of the 
resolved and the unresolved 15~$\mu$m members, respectively.  
The 15~$\mu$m fluxes of the 15~$\mu$m
members are indicated by three different sizes of the symbols 
(brighter objects are shown in larger symbols).

A close-up map of the central $5' \times  5'$ region is shown
in Fig.~\ref{fig:map_IR}(b).
It is worth noting that a $5' \times 5'$ field corresponds to
the field of view of a single pointing with Spitzer MIPS, 
and our total spatial coverage with AKARI is $\sim 8$ times 
wider than that.
Our data is deepest at this cluster central region 
because this region is covered by all the F1, F2 and F12 fields.
Therefore, we here plot the 15~$\mu$m members all the way to 
the faintest end
at 15~$\mu$m ($\gsim$ 3.5$\sigma$ detection).   
However, we rarely see 15~$\mu$m members in the very central 
region of the cluster (i.e. $\lsim 1'$ from the centre),
although the number density of the phot-$z$ cluster members
is the highest in this very central region. 
In contrast, the outer regions just outside this zone of avoidance 
of the 15~$\mu$m members, many 15~$\mu$m members are detected.
In particular, they are distributed preferentially at the 
region $\sim 1'$ north-east and south-west away from the centre. 
These regions are located right
on the large scale NE--SW filament penetrating the cluster
core and hosting two distinct groups (see Fig.~\ref{fig:map_IR}a).

\begin{figure}
   \begin{center}
    \leavevmode
    \vspace{-1cm}
    \begin{center}
    \leavevmode
    \rotatebox{0}{\includegraphics[width=8.5cm,height=8.5cm]{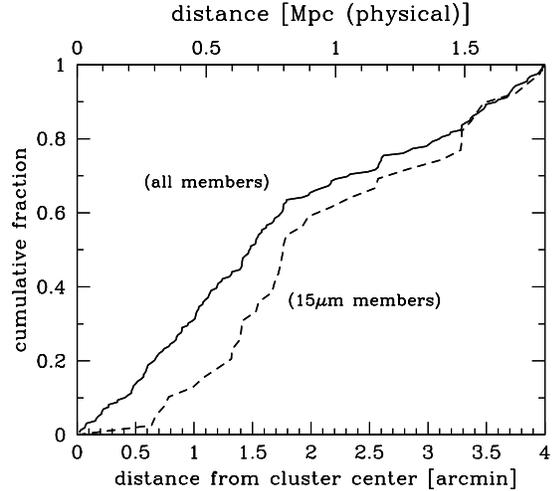}}
    \end{center} 
   \end{center} 
   \caption{ Cumulative fractions of all the phot-$z$ members 
(solid line) and all the 15~$\mu$m members (dashed line)
as a function of the distance from the cluster centre.
The 15~$\mu$m members are much less concentrated in the central 
$1'$ region, while they dramatically increase within 
the $\sim 1.5-2.0$ arcmin ring from the centre. 
These plots are made for galaxies with 
$z'<22.5$ only, because most of the 15~$\mu$m members are detected 
from such optically bright galaxies 
(see text in Section~\ref{subsec:15um_members_on_cmd}).  
}
\label{fig:rad_profile}
 \end{figure}

To show it more quantitatively, we plot the cumulative radial profiles
of the member galaxies as a function of distance from the cluster 
centre out to $4'$ in Fig.~\ref{fig:rad_profile}.
Fig.~\ref{fig:rad_profile} clearly shows that 
the 15~$\mu$m members are less concentrated in the $\lsim 1'$ 
region of the cluster than the general cluster member
galaxies.  
From the slope of the curve for the 15~$\mu$m members, 
we see that many 15~$\mu$m members are located in the radius 
range of $\sim$1.5--2.0
arcmin from the centre, which in fact corresponds to the region 
just around the cluster centre (see also Fig. \ref{fig:map_IR}b).
A Kolmogorov-Smirnov test on these two subsamples shows that
the probability that these two are from the same parent is
only $\sim$ 3.5 \%. 
A more detailed analysis of the distribution of the 15~$\mu$m member
galaxies will be carried out in Section 
\ref{subsec:enhancement_of_dusty_SF}.

%%%%%%%%%%%%%%%%%%%%%%%%%%%%%%%%%%%%%%%%%%%%%%%
% Enhancement of dusty SF activity in the cluster outskirts
%%%%%%%%%%%%%%%%%%%%%%%%%%%%%%%%%%%%%%%%%%%%%%%
\subsection{Enhancement of dusty star formation activity in the cluster outskirts}
\label{subsec:enhancement_of_dusty_SF}

\begin{figure}
   \begin{center}
    \leavevmode
     \rotatebox{0}{\includegraphics[width=8.5cm,height=8.5cm]{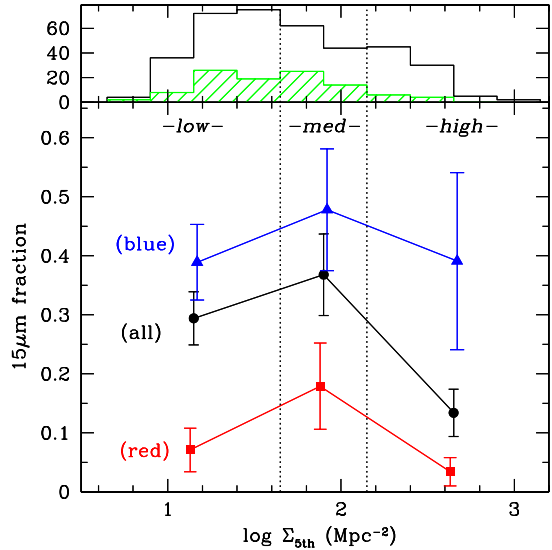}}
   \end{center}
\caption{ Fractions of 15~$\mu$m member galaxies with 
$f(15 \mu m) \ge 67 \mu \textrm{Jy}$ as a function of the environment.  
The trend that the fraction is highest in the 
medium-density environment (especially for red galaxies)
is to be noted.  The open and hatched histogram in the upper 
panel represent the distribution of $\log \Sigma _5$ for all 
the phot-$z$ members with $z' \le 22.5$ and for 15~$\mu$m members 
with $z' \le 22.5 $, respectively.  The vertical dotted lines
at $\log \Sigma _5 =$ 1.65 and 2.15 represent the definitions
of low, medium and high density environments. }
\label{fig:15um_fraction}
\end{figure}

For each environment (i.e. low, medium and high density), 
we calculate the fraction of the 15~$\mu$m
members in all the member galaxies with $z' \le 22.5$,
including both resolved and unresolved 15~$\mu$m members. 
We show in Fig. \ref{fig:15um_fraction} the fractions 
of the 15~$\mu$m members with $f(15 \mu \textrm{m}) \ge
67 \mu \textrm{Jy}$ for red, blue and red+blue 
cluster member galaxies as a function of the environment. 
The boundary between red and blue is set as 
$$ (R-z') = 2.51 - 0.049 \times z' ,$$
which is defined as 0.2 mag bluer than the best-fitted
CMR shown in \cite{koy07} (see also Fig.~\ref{fig:colmag_all}a). 
We only used the galaxies with $z' \le 22.5$ because the number of
15~$\mu$m members with $z' \ge 22.5$ is very small due to the depth 
of the 15~$\mu$m image (see Fig.~\ref{fig:colmag_all}a). 

We show in Fig.~\ref{fig:15um_fraction} that  
the 15~$\mu$m fraction for all galaxies decreases
dramatically in the high-density environment compared
with low and medium-density environments.  This represents
the very low star forming activity in the cluster core. 
On the other hand, we find that the 15~$\mu$m fraction is 
still high in the medium-density environment compared 
with the low-density fields, although we showed in 
Section~\ref{subsec:optical_colour_transition} that 
the galaxy colours starts to sharply change in the
medium-density environments. 
Although the statistics is not very good, 
the fraction seems to be higher in the medium-density 
environment than in the low and high-density environments.
Such a trend is likely to be strong for the red galaxies.
This may indicate that in the medium-density environments
dusty red galaxies are preferentially produced and/or 
the star forming activity is once enhanced for some galaxies
(see below).

In this calculation, we included all the 15~$\mu$m members 
down to the limit of $f(15\mu \textrm{m}) = 67 \mu$Jy
(i.e. $5.0\sigma$ in the deep region and $3.5 \sigma$ in the
shallow region),
but we found that this trend does not change if we limit 
the sample to relatively brighter 15~$\mu$m members 
(e.g. $f(15\mu \textrm{m}) \ge 100 \mu$Jy which corresponds 
to $7.5\sigma$ in the deep region and $5.2 \sigma$
in the shallow region).
Also, as noted above, we included both resolved and 
unresolved 15~$\mu$m cluster members in this calculation.  
If we use only the resolved 15~$\mu$m cluster members,
the significance of the high fraction of optically 
red 15~$\mu$m members in the medium-density environment
becomes even higher.  

\begin{figure}
   \begin{center}
    \leavevmode
    \rotatebox{0}{\includegraphics[width=8.5cm,height=8.5cm]{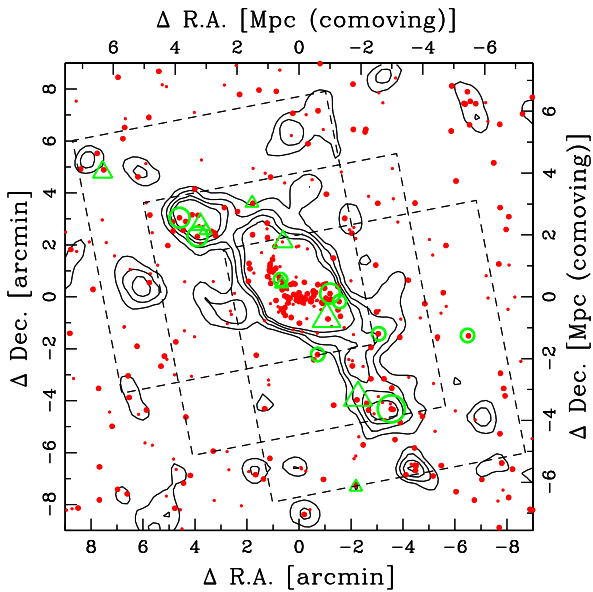}}
   \end{center} 
   \caption{ The spatial distribution of the optically red 15~$\mu$m 
  cluster members (open circles and triangles) on all optically red 
  cluster members. 
  The size of the open symbols indicates the 15~$\mu$m flux as in
  the previous figures (e.g. Fig.~\ref{fig:map_IR}).  
  The large and small filled dots represent
  the $z'\le 22.5$ and $z' > 22.5$ optically red cluster members,
  respectively.
  We can see that the distribution
  of optically red 15~$\mu$m members is aligned in the direction of the
  filament, and matches well with the distribution of the galaxies in
  the medium-density environment (groups/filaments). }
\label{fig:map_dusty_SF}
 \end{figure}

We now focus on the optically red 15~$\mu$m members 
(i.e.\ dusty red galaxies) that we have noted in
Section~\ref{subsec:15um_members_on_cmd}.
We find 19 such galaxies in the entire field, and
they are detected at significantly larger fraction in the 
medium-density environment compared to the
low and high density environments (more than $\sim 2$ times larger).
Detection in 15~$\mu$m of the optically red galaxies should mean
that these galaxies still have on-going star formation activity,
although their $R-z'$ colours are as red as passively evolving galaxies. 
As described in Section~\ref{subsec:15um_members_on_cmd}, these 
galaxies are strong candidates for star forming galaxies
heavily attenuated by dust. 
Our results therefore suggest that the dusty star formation
activity is induced by some mechanisms which are effective
in the medium-density environment.
We plot the spatial distribution of these optically red 15~$\mu$m
cluster member galaxies in Fig.~\ref{fig:map_dusty_SF} on top
of all the optically red phot-$z$ members. 
It is quite noticeable that almost all dusty red galaxies
are distributed exclusively along the filament including groups 
traced by the optical members, and in fact, this distribution 
is very similar to that of the member galaxies in the 
medium-density environment 
(see Fig.~\ref{fig:colour_density}b).  
Although there are a few dusty red galaxies in 
the low and high density environments as well (Fig. \ref{fig:15um_fraction}), 
almost all of them are located immediately outside of the filaments.

%---------------------------------------------------------------
%---------------------------------------------------------------
%----------------------------------------------------------------------
%-------------------------------------------------------------------------
%-------------------------------------------------------------------------
%-------------------------------------------------------------------------

\begin{figure*}
   \begin{center}
    \leavevmode
    \epsfxsize 0.48\hsize
    \epsfbox{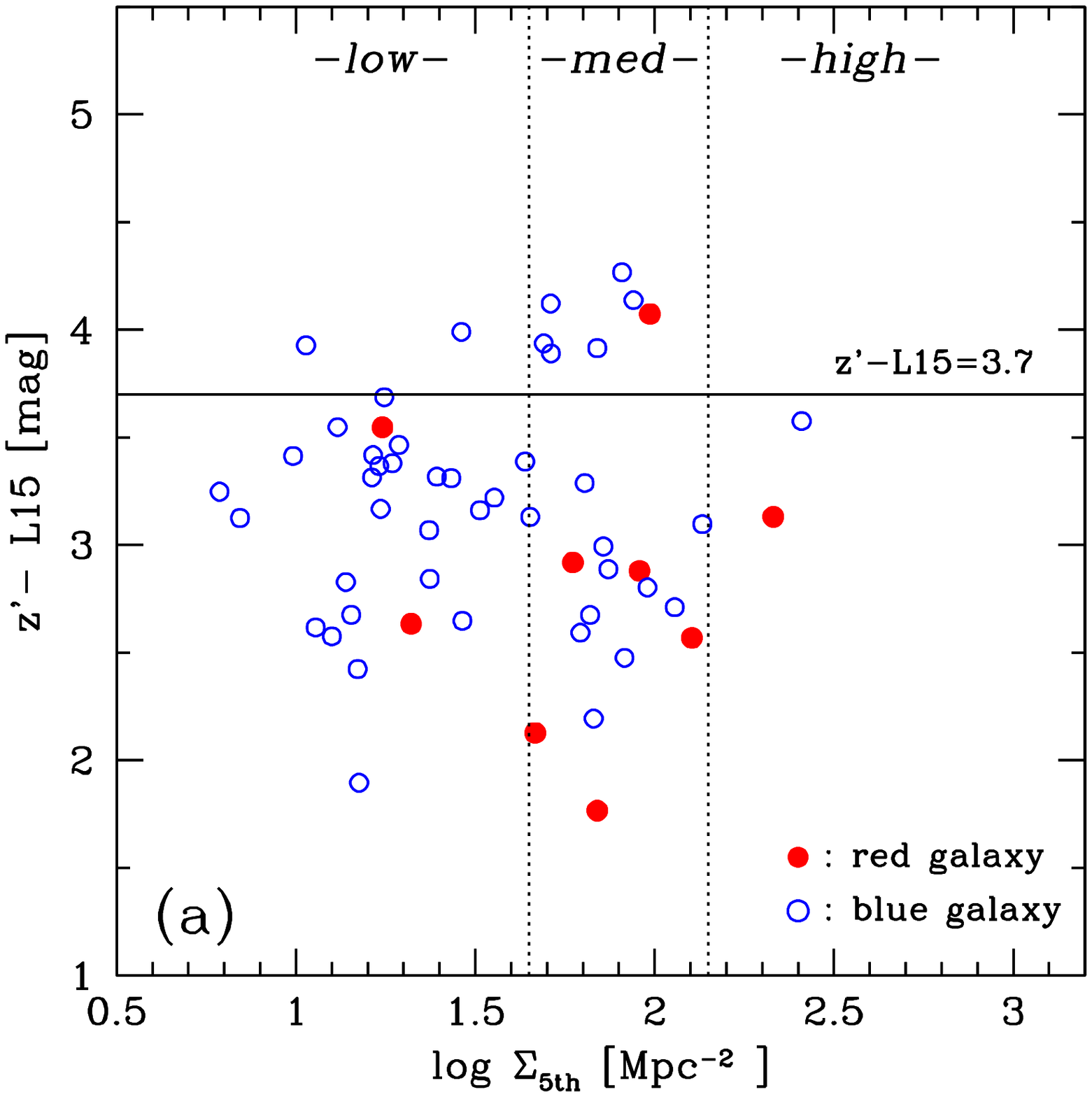}
    \epsfxsize 0.48\hsize
    \epsfbox{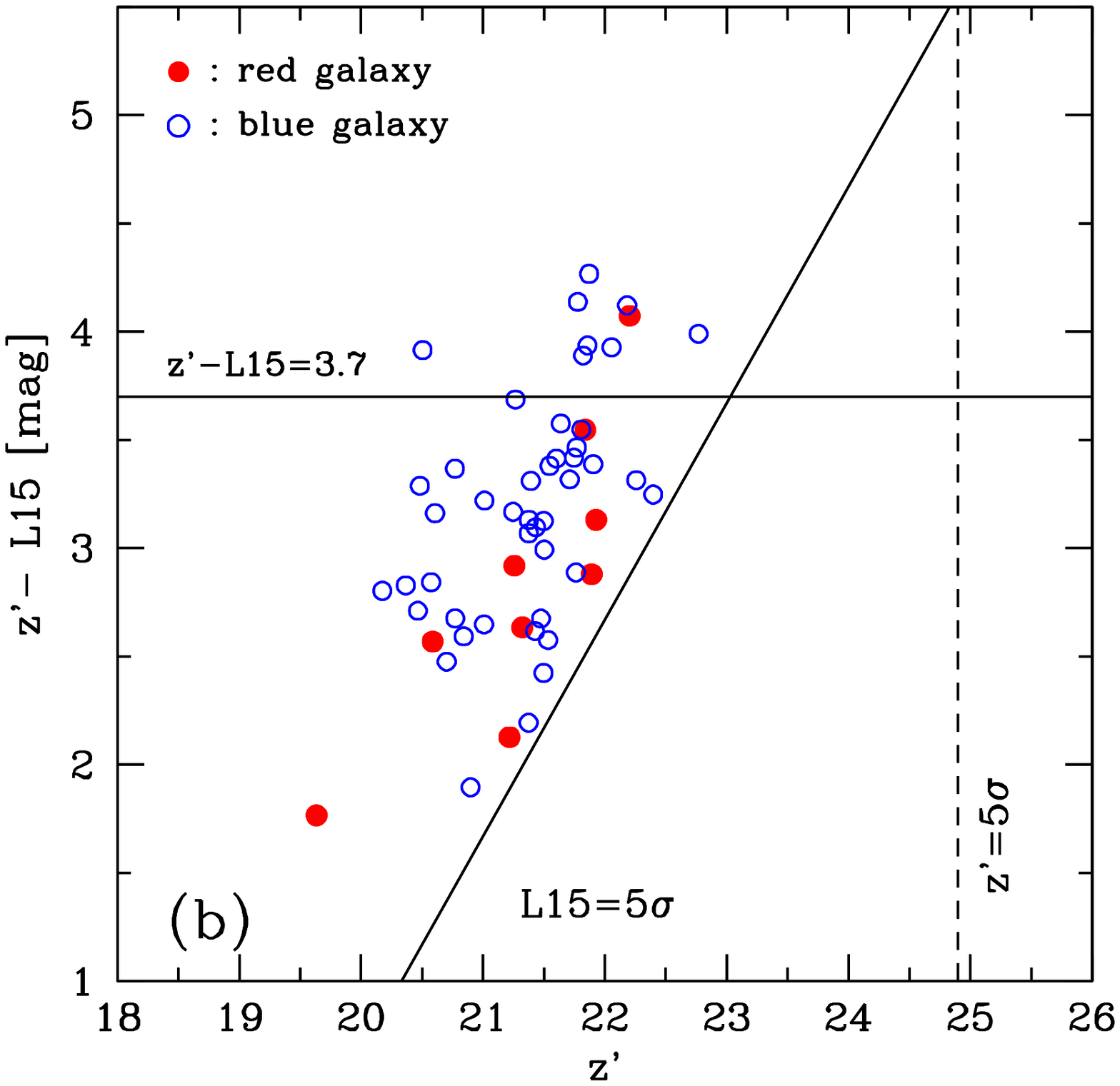}
   \end{center}
   \caption{ $z' - \textrm{L15}$ colours as a function of the local
 density (a), and the $z'$-band magnitude (b). 
 Filled and open symbols indicate optically red and blue 
 resolved 15~$\mu$m cluster members, respectively.  Galaxies above the 
 horizontal solid line at $z' - \textrm{L15} =3.7$ are defined as
 ``high SSFR galaxies''.  Note that these high SSFR 
 galaxies are concentrated in the medium-density environment 
 and concentrated at around $21.5 \lsim z' \lsim 22$.  The vertical
dotted lines in panel(a) represent the definitions of the environment.
The vertical dashed line and slanted solid line in panel(b) show the
5$\sigma$ limiting magnitudes in $z'$ and L15, respectively.}
\label{fig:z_L15}
\end{figure*}

To study the properties of the 15~$\mu$m members in the medium-density
environment further, we investigate the $z' - \textrm{L15}$ colours
for the 15~$\mu$m members.  
The $z'$-band magnitudes approximate the 
stellar mass of galaxies ($\sim 5000$ \AA { } in the 
rest frame for $z\sim 0.8$ galaxies) and the 15~$\mu$m magnitude
approximates the star formation rate. 
Therefore, $z'- \textrm{L15}$ colour approximately 
corresponds to ``specific star formation rate (SSFR)'' 
which is defined as star formation rate per unit stellar mass of
galaxies.  Since SSFR has a unit of inverse of time, it can be regarded
as a timescale of star formation activity (i.e. galaxies
which have high SSFR are considered to have short star formation 
timescales).
In Fig.~\ref{fig:z_L15}, we plot the $z' - \textrm{L15}$
colours for the resolved 15~$\mu$m members
as functions of local density and $z'$-band magnitude.

%-------------------------------------------------------------------------
%-------------------------------------------------------------------------

\begin{figure}
   \begin{center}
    \leavevmode
    \rotatebox{0}{\includegraphics[width=8.5cm,height=8.5cm]{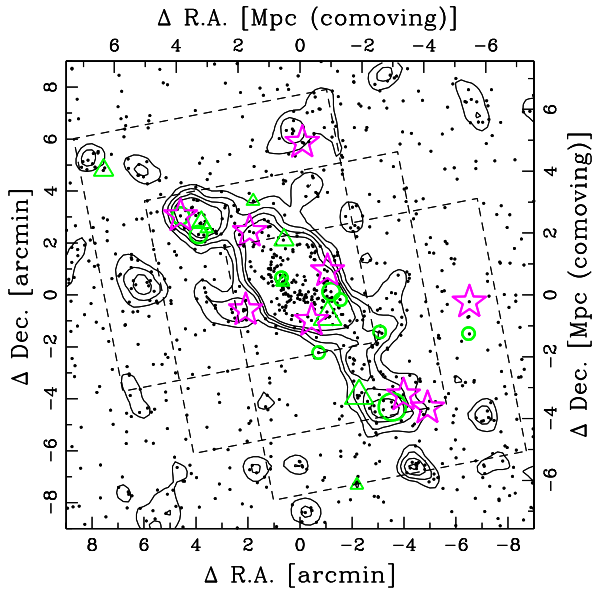}}
   \end{center} 
   \caption{ The distribution of the high SSFR galaxies (open stars).
Small dots indicate all the phot-$z$ member galaxies.  The high SSFR 
galaxies are also distributed preferentially in the
medium-density environment (i.e. groups/filaments). 
Open circles and triangles represent the ``dusty 
red galaxies'', same as in Fig.~\ref{fig:map_dusty_SF}.}
\label{fig:map_high_SSFR}
 \end{figure}

We notice that 9 out of 54 galaxies 
plotted in Fig.~\ref{fig:z_L15} have 
large colour indices of $z' - \textrm{L15} > 3.7$.
We can consider that these galaxies are forming stars vigorously and
efficiently for their stellar masses.  This boundary roughly corresponds
to $1/\textrm{SSFR} \sim 0.25$ Gyr for $z'=22$ mag galaxies. 
This calculation is based on the stellar mass of a galaxy calculated 
from the $z'$-band magnitude using the disk galaxy model 
in \cite{kod99} and the SFR calculated from the 15~$\mu$m magnitude. 
We hereafter call these galaxies ``high SSFR galaxies''. 
Interestingly, many of the high SSFR galaxies also 
tend to live in the medium-density environment
(see Fig. \ref{fig:z_L15}a).  
We note that 7 out of 23 ($\sim$ 30\%) 15~$\mu$m members satisfy 
the high SSFR criterion in the medium-density environment, 
while only 2 out of 29 ($\sim$7\%) galaxies in the low-density 
environments are high SSFR galaxies, and there is no such galaxy 
in the high-density environments.  
In Fig. \ref{fig:z_L15}(b), 
we can see that the $z'$-band magnitudes of such high SSFR 
galaxies tend to be fainter than $z' \sim 21.5$.  
We show the spatial distribution of such high SSFR galaxies 
in Fig.~\ref{fig:map_high_SSFR} with large open stars.
The high SSFR galaxies also tend to be seen in or just around 
the filament, although their distribution may be slightly more extended 
than that of the dusty red star forming galaxies 
(i.e. open circles and triangles in Fig.~\ref{fig:map_high_SSFR}).

We note that in Section~ \ref{subsec:derive_sfr} we found
five possible AGN candidates, at which three have X-ray 
detections and the other two have red N3$-$S7 colours.
We find none of these five sources satisfy the high
SSFR criterion of $z'-\textrm{L15} > 3.7$.  
We find two optically red sources in these five AGN 
candidates (one in the low-density environment and the other 
in the medium-density environment).  
Therefore, we find that these AGN candidates do not strongly
affect our results on the high SSFR galaxies and the dusty red 
galaxies.
We stress here that even if these galaxies are really 
associated with pure AGNs and if we exclude these galaxies 
from our sample, our conclusion is not changed. 

It is found that both the ``dusty red galaxies''
and the ``high SSFR galaxies'' prefer to live in the medium-density
environment.  Since this environment is the one where colour distribution
of galaxies start to change drastically from blue to red (see Section~
\ref{subsec:optical_colour_transition}), it is reasonable to
think that these galaxies are related to such colour transition
of galaxies, and they may be
at the transient phase from blue active galaxies to red passive galaxies
during the course of hierarchical assembly to higher density regions.

%---------------------------------------------------------------
%------------------------------------------------------------------------

%%%%%%%%%%%%%%%%%%%%%%%%%%%%%%%%%%%%%%%%%%%%%%%%%%%%%%%%%%%%%%%%%%%%%%%%
% DISCUSSION
%%%%%%%%%%%%%%%%%%%%%%%%%%%%%%%%%%%%%%%%%%%%%%%%%%%%%%%%%%%%%%%%%%%%%%%%
\section{Discussion}
\label{sec:discussion}

% comparison with the literatures %

In this section, we discuss our results in broader context
of environmental dependence of galaxy evolution in clusters
by comparing with previous works in the literature.
Since our study is the first attempt to investigate around 
a $z\sim 1$ cluster with such a wide field coverage in MIR
observation, we cannot directly compare our results with others
for the region far out from the cluster centre.  As for
the cluster central regions, two $z \sim 0.8$ clusters 
(RXJ0152 and MS1054, both at $z=0.83$) are studied 
recently with Spitzer MIPS by \cite{mar07} 
and \cite{bai07}. 
\cite{mar07} found that MIR cluster members are distributed 
outside the two main clumps traced by X-ray emissions in 
the RXJ0152 cluster.  Similarly, \cite{bai07} showed that
there are few IR galaxies in the high-density regions of the
cluster.  For the RXJ1716 cluster, we find in Section.~4.3 
that the 15~$\mu$m member galaxies are distributed avoiding 
the central part of the cluster, which is qualitatively 
consistent with the results of \cite{mar07} and \cite{bai07}.  
The absence of luminous infrared galaxies in the cluster 
centre is also reported for some lower-$z$ clusters, including 
local Coma and Virgo clusters by ISOCAM (see review 
by \citealt{met05}).
Therefore, we have confirmed the low star formation activity 
in the high-density environment such as cluster centres
with the MIR observations to $z \sim 0.8$. 
It is natural to think that
the red optical colours of the galaxies in cluster centres
are primarily due to the lack of star formation activity and 
not due to the reddening by dust.

% about high SFR in filaments %

We found that MIR bright cluster members are detected
relatively far out from the cluster centre, especially 
in group/filament environment.
Very recently, \cite{fad08} conducted a very wide-field MIR observation
with Spitzer MIPS which covers two filaments around 
the Abell 1763 cluster ($z = 0.23$), and found that the 
fraction of starburst galaxies is more than twice larger 
in the filaments than 
in the inner region or outer fields of the clusters.
The enhancement of star forming activity in filaments 
is also suggested in nearby clusters in the optical 
studies (e.g. \citealt{por07}; \citealt{por08}).    
Our result for the cluster at $z=0.81$ is qualitatively
consistent with these studies in the sense that 
star forming activity is enhanced in the intermediate
density environment between low-density general field and
the high-density cluster core.  From the field study
at $z \sim 1$, it is reported that the environment of LIRGs/ULIRGs 
is denser than that of field galaxies (e.g.\ \citealt{mar08}).
\cite{elb07} studied a structure at $z\sim 1$ in the
GOODS-North field with Spitzer MIPS, and showed that galaxies
with strong star formation are preferentially seen in group centres.
They suggested that, at $z \sim 1$, SFR of individual galaxies
increases with increasing density up to a certain critical
density and it decreases again at higher density 
(see also \citealt{coo07}).  For distant clusters,
\cite{pog08} recently suggest a possible peak in the SFR-density 
relation based on the \oii { }line study of EDisCS clusters
at $z=0.4 - 0.8$.
These studies are also qualitatively consistent with our results, 
although the structures investigated in \cite{elb07} would be 
much poorer systems than the RXJ1716 cluster, 
judging from their weak X-ray detection.

% about dusty red galaxy  %

We now focus on the optically red star forming 
galaxies (i.e.\ ``dusty red galaxies''). 
Dusty red star forming galaxies in clusters were studied 
in some previous works. 
For example, \cite{coi05} studied the CL0024+1654 cluster
at $z \sim 0.39$ with ISOCAM, and found that about half of the
MIR sources in the cluster reside on the red-sequence. 
These dusty galaxies are not concentrated in high-density
regions.
Based on the COMBO-17 data for the A901/902 clusters at
$z \sim 0.17$, \cite{wol05} also showed that more than
1/3 of the red-sequence galaxies have dusty red SEDs.
They also found that these dusty red galaxies prefer
medium-density outskirts of the clusters and they are rare
in low or high density environments.  Since the redshifts
of these clusters are lower than that of RXJ1716 ($z = 0.81$),
we cannot conclude that these are the same populations
as our dusty red galaxies.
However, judging from their optically red colours while having
signs of on-going star formation activity, we expect that our dusty
red galaxies are similar counterparts of the dusty
populations seen in the two low-$z$ clusters studied
by \cite{coi05} and \cite{wol05}.
As we showed in Fig.~\ref{fig:map_dusty_SF}, our dusty 
red galaxies are preferentially located in the
medium-density environment.  It is interesting to note that these 
optically red star forming galaxies are common in the outskirt
of clusters at all redshifts through $0.1 \lsim z \lsim 0.8$,
although the number of cluster sample is very limited.

% Finally physical process for the transformation 

\begin{figure}
   \begin{center}
    \leavevmode
    \rotatebox{0}{\includegraphics[width=8.5cm,height=5.5cm]{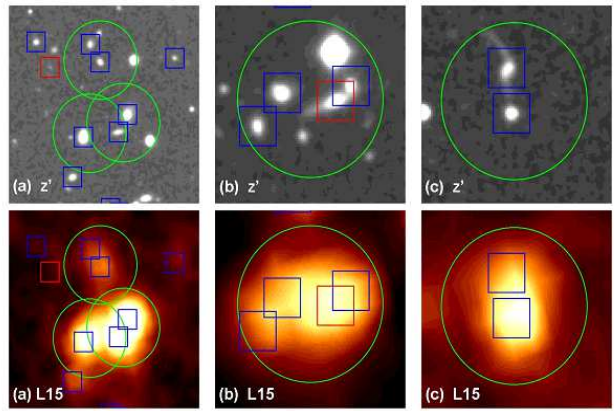}}
   \end{center} 
   \caption{ Examples of the optical counterparts for 15~$\mu$m members 
   (upper panels) and their 15~$\mu$m image (lower panels).  
   Circles show the positions of 15~$\mu$m members and
   their radius are set to 8$''$ in all the panels.
   Small squares represent the positions of phot-$z$
   selected member galaxies and the size of the squares are 
   4$'' \times$ 4$''$.  (a) is a candidate for an in-falling group.
   (b) and (c) are the prominent cases of the interacting systems
   with 15~$\mu$m emissions.
\label{fig:interesting_obj}}
 \end{figure}

Finally, we discuss the physical mechanisms that can play major roles
in galaxy truncation of star forming activity.
\cite{mar07} found that MIR members in the 
RXJ0152 cluster seem to be mostly associated with the 
in-falling late-type galaxies classified by \cite{bla06}.  
They suggest that a burst of star formation
can occur during the galaxy in-fall process.  Since we do not
have spectroscopic data for our 15~$\mu$m cluster members,
we cannot be sure at this stage that our 15~$\mu$m members
are physically associated to the in-falling galaxies, but it is likely
that at least some of our 15~$\mu$m cluster members in the
RXJ1716 cluster are indeed the galaxies that are in-falling
along the filament to the main body of the cluster by gravity.
In fact, we have some good candidates of in-falling groups
far away from the cluster centre located at
around (6.\hspace{-2pt}$'$0, 0.\hspace{-2pt}$'$0), (0.\hspace{-2pt}$'$0, 
6.\hspace{-2pt}$'$0) and ($-$2.\hspace{-2pt}$'$0, 2.\hspace{-2pt}$'$0) in
Fig.\ \ref{fig:map_IR}(a) as the iso-density contours clearly show.
We show in Fig.~\ref{fig:interesting_obj}(a) an example of the
candidates of an in-falling group located at
($-$0.\hspace{-2pt}$'$5, 1.\hspace{-2pt}$'$5) 
in Fig.~\ref{fig:map_IR}(b) which is $\sim$ 1 Mpc away from the 
cluster centre.
We need spectroscopic confirmation of the physical
association of these systems to the cluster.

We have revealed that the 
15~$ \mu$m members are found far out from the cluster 
core especially in the medium-density environment such 
as groups or filaments where the optical colour distribution
strongly changes.  Our results suggest that many galaxies 
which entered medium-density environment from the low-density field
experience starburst, and these galaxies are observed 
as dusty star forming galaxies and/or high SSFR galaxies 
around the RXJ1716 cluster.
Moreover, it would be natural to consider that the burst of 
star formation is linked to
the colour and morphological changes of galaxies and truncation
of star forming activity. 
The most likely physical mechanism at work
in these medium density regions which also involves starburst
would be galaxy-galaxy interaction or mergers (e.g.\ \citealt{hop08}). 
We have two good examples of 15~$\mu$m members which show 
prominent interacting features (Fig.~\ref{fig:interesting_obj}b,c).  
The strong MIR emissions from these systems must be produced 
in the process of the galaxy interaction.  
These findings may suggest a link between the
dusty star forming activity of galaxies and galaxy-galaxy 
interaction.  
However, it is difficult to firmly conclude that the majority of our 
15~$\mu$m galaxies are activated via galaxy-galaxy interactions
or mergers.  The seeing size of our ground-based optical image
is $\sim$0.\hspace{-2pt}$''$7 and it is not sufficient to 
determine the morphology for all of our 15~$\mu$m galaxies 
and to find any interacting signatures.  
We should also admit that spectroscopic confirmation of 
membership of the phot-$z$ members especially for the 15~$\mu$m
members is crucial to draw any firm conclusions.
Since our study reaches far out from the cluster core, 
it would have more contaminations than other studies limited 
in the cluster cores.
The surface number density of the contaminant galaxies with 
$z' \le 22.5$ mag is estimated to be $\sim 0.93$ arcmin$^{-2}$, 
using the galaxies in the control fields defined in our Suprime-Cam field 
(see Fig.~3 of \citealt{koy07}).  
A rough estimate shows that, among all the phot-$z$ members 
with $z'\le 22.5$ mag, $\sim$10\% and $\sim$25\% galaxies 
can be contaminant galaxies in the cluster region (i.e. high-density region) 
and the group region (i.e. medium-density region),
respectively (see Fig.~4 of \citealt{koy07} for the
definitions of the cluster and the groups).  
Furthermore, we should keep in mind that the properties and/or fraction
of the MIR galaxies may be different from cluster to cluster,
as suggested in \cite{gea06}, even if we investigate 
clusters nearly at the same redshifts.
Wide-field MIR observations on a larger sample of distant clusters
is clearly required to obtain more general picture of the 
environmental effects on galaxy evolution.

%------------------------------------------------------------------
%-----------------------------------------------------------------

%%%%%%%%%%%%%%%%%%%%%%%%%%%%%%%%%%%%%%%%%%%%%%%%%%%%%%%%%%%%%%%%%%%%%%
% SUMMARY AND CONCLUSION
%%%%%%%%%%%%%%%%%%%%%%%%%%%%%%%%%%%%%%%%%%%%%%%%%%%%%%%%%%%%%%%%%%%%%%
\section{Summary and Conclusions}
\label{sec:summary}
We have performed a wide-field and multi-wavelength
optical and infrared study of the distant galaxy cluster
RXJ1716.4+6708 (RXJ1716) at $z=0.81$.
A unique wide field coverage both in optical and infrared
has enabled us to classify galaxies into three environmental bins,
namely, high-density regions (cluster core),
medium-density regions (cluster outskirts, groups, filament),
and low-density regions (field), and has thus allowed us to investigate
galaxy properties as a function of environment along the structures
in and around the distant cluster.

We find many of the 15~$\mu$m cluster members show intermediate
optical colours between the red sequence and the blue cloud.
This may indicate that these 15~$\mu$m cluster members are
actively star-forming galaxies but attenuated by dust 
hence showing intermediate optical colours, 
although relatively fewer detection at 15~$\mu$m 
of blue galaxies may be partly because they tend to be 
optically faint.    

We quantified the environment around the cluster
using the local projected number density of cluster
member galaxies, and confirmed that the optical colour
distribution starts to dramatically change
at the ``medium'' density environment that corresponds
to groups and/or filaments.  We showed that the 15~$\mu$m
members are very rare in the high-density cluster centre. This is 
probably due to the low star forming activity in such regions.
However, interestingly, the fraction of the 15~$\mu$m-detected 
cluster members in the medium-density environment is as high 
as in the low-density fields, despite the fact that optical
colours of galaxies start to dramatically change from blue to red in the 
medium-density environment.  Although the statistic is not very good, 
the fraction is slightly higher in the medium-density 
environments even compared with the low-density fields.

We also find that dusty red galaxies
(optically red 15~$\mu$m cluster members) and the galaxies 
with high specific star formation rates (red in $z'-$ 15~$\mu$m
colour) are both concentrated in the medium-density environment.
These results may suggest that the star formation activity
in galaxies is once enhanced by some physical processes which
are effective in group/filament environment (e.g.
galaxy-galaxy interaction), before their star forming 
activity is eventually truncated and they move on to the red sequence.

We stress that all these new findings are based on 
the widest field MIR observation of a $z \sim 0.8$ cluster so far. 
There is no other study that covers such a wide field 
around clusters in MIR at $z \gsim 0.8$.  Since our study 
is a case study for just one cluster at $z = 0.81$,
we are desperately in need for a larger sample of distant 
clusters viewed in the infrared regime and at the same time 
covering a wide field of view so that we can witness the 
galaxy truncation in action in the in-fall regions
of distant clusters along the filaments.
This is essential in order to confirm this interesting trend and
to obtain a general view of galaxy evolution.

%%%%%%%%%%%%%%%%%%%%%%%%%%%%%%%%%%%%%%%%%%%%%%%%%%%%%%%%%%%%%%%%%%%%%%
% ACKNOWLEDGEMENT
%%%%%%%%%%%%%%%%%%%%%%%%%%%%%%%%%%%%%%%%%%%%%%%%%%%%%%%%%%%%%%%%%%%%%%
\section*{Acknowledgement}

We thank the anonymous referee for the careful reading of 
the paper and for the helpful suggestions, which improved the
paper.
We thank Dr. Tomotsugu Goto and Dr. Kimiaki Kawara for 
their helpful suggestions and discussions.  
We are also grateful to Dr. Takashi
Onaka for advice on the AKARI data treatment.
This work was financially supported in part by a Grant-in-Aid for the
Scientific Research (Nos.\, 15740126; 18684004) by the Japanese 
Ministry of Education, Culture, Sports and Science. 
The infrared data that we used in this study is based on 
observations with AKARI, a JAXA project 
with the participation with ESA.  
The optical data is collected at the Subaru Telescope, which 
is operated by the National Astronomical Observatory of Japan. 
Y.K. and T.T. acknowledges support from the Japan Society for the Promotion 
of Science (JSPS) through JSPS research fellowships for Young Scientists.
M.I. was supported by Creative Research Initiatives grant 
R16-2008-015-01000-0 of MOST/KOSEF. H.M.L was supported by 
the Space Science Development Program from KASI.

%------------------------------------------------------------------

%------------------------------------------------------------------
\end{document}